\documentclass[aps,prl,twocolumn,superscriptaddress,english]{revtex4-2}

\usepackage[T1]{fontenc}
\usepackage{babel}
\usepackage{amsmath}
\usepackage{amssymb}
\usepackage{wasysym}
\usepackage{graphicx}
\usepackage{lipsum}
\usepackage{xcolor}
\usepackage{braket}
\usepackage{booktabs}
\usepackage{multirow}
\usepackage{makecell}
\usepackage{xcolor}

\usepackage[linktocpage=true,
  colorlinks=true, 
  pdfborder={0 0 0},
  linkcolor=blue,
  citecolor=red,
  filecolor=yellow,
  urlcolor=blue,
  bookmarks,
  pdfauthor={},
]{hyperref}

\newcommand{\tc}{$T_\text{c}$}

\newcommand{\ep}{\textit{e-ph}~}

\begin{document}

\title{LaBH$_{8}$: the first high-\tc{} low-pressure superhydride}

\author{Simone Di Cataldo} \email{simone.dicataldo@uniroma1.it}
\affiliation{Institute of Theoretical and Computational Physics, Graz University of Technology, NAWI Graz, 8010 Graz, Austria}
\affiliation{Dipartimento di Fisica, Sapienza Universit\`a di Roma, 00185 Roma, Italy} 
\author{Christoph Heil}
\affiliation{Institute of Theoretical and Computational Physics, Graz University of Technology, NAWI Graz, 8010 Graz, Austria}
\author{Wolfgang von der Linden}
\affiliation{Institute of Theoretical and Computational Physics, Graz University of Technology, NAWI Graz, 8010 Graz, Austria}
\author{Lilia Boeri} \email{lilia.boeri@uniroma1.it}
\affiliation{Dipartimento di Fisica, Sapienza Universit\`a di Roma, 00185 Roma, Italy} 

%% ABSTRACT: MAX 600 WORDS
\begin{abstract}
In the last five years a large number of new high-temperature superconductors have been predicted and experimentally discovered among hydrogen-rich crystals, at pressures which are way too high to meet any practical application. 
In this work, we report the computational prediction of a hydride superconductor, LaBH$_{8}$, with a \tc{} of 126 K at a pressure of 50 GPa, thermodynamically stable above 100 GPa, and dynamically stable down to 40 GPa, an unprecedentedly low pressure for high-\tc{} hydrides.
LaBH$_{8}$ can be seen as a ternary sodalite-like hydride, in which a metallic hydrogen sublattice is stabilized by the chemical pressure exerted by the guest elements. The combination of two elements with different atomic sizes in LaBH$_{8}$ realizes a more efficient packing of atoms than in binary sodalite hydrides. A suitable choice of elements may be exploited to further reduce the stabilization pressure to ambient conditions.
\end{abstract}

%% MAIN TEXT: MAX 3750 WORDS
%% WORD COUNT INCLUDES:
%%% Text in the body of the article
%%% Text in figure and table captions 
%%% Text in footnotes and endnotes

%% WORD COUNT DOES NOT INCLUDE:
%%% Title, authors, affiliation
%%% Abstract
%%% References
%%% Acknowledgments

%% To estimate the word equivalent for figures use the figure’s aspect ratio (width / height). The estimate is [(150 / aspect ratio) + 20 words] for single-column figures, and [300 / (0.5 * aspect ratio)] + 40 words for double-column figures.

%% Tables: the word equivalent for tables is 13 words plus 6.5 words per line for single-column tables. Double-column tables count as 26 words plus 13 words per line.

\maketitle
%% INTRODUCTION
The discovery of high-temperature superconductivity (HTSC) at 203 K in sulfur hydride at a pressure of 200 GPa rekindled the dream of achieving
room-temperature superconductivity \cite{Eremets_NatPhys_2016_SH3, Eremets_Nature_2015_SH3,  Duan_SciRep_2014_SH}, triggering a \textit{hydride rush} \cite{Eremets_arXiv_2019_YH6, Oganov_arXiv_2019_YH6, Oganov_Science_2018_UH, Oganov_MatToday_2020_ThH, Eremets_arXiv_PH3_2015,Eremets_Nature_2019_LaH, Hemley_PRL_2019_LaH, Boeri_PhysRep_2020_review}
which culminated in the report of superconductivity with a critical temperature (T$_c$) of 
287 K (15 $^{\circ}$C) at 267 GPa in a carbonaceous sulfur hydride \cite{Dias_Nature_2020_CSH}.
The first discovery of a room-temperature superconductor set a major 
milestone in the history of condensed matter physics,
but the exceptional pressure required to stabilize the superconducting phase
thwarts any practical application.

Obviously, the next challenge for materials research is to find materials
exhibiting comparable superconducting properties at, or close to, ambient pressures.
Also in this case hydrides, which realize the requirements for conventional
HTSC, are a promising hunting ground.

In the five years following the SH$_3$ discovery,
all possible combinations of $X$H$_n$ \textit{binary} hydrides have been computationally explored in an effort to achieve room-temperature superconductivity; these studies revealed that the formation, stability and superconducting properties of these high-pressure (HP) hydrides strongly depend on the ionic size, electronegativity and electronic configuration of the $X$ elements. High-T$_c$ superconductors are found either among \textit{covalent}
hydrides, in which $X$ and H form directional bonds driven metallic by pressure,~\cite{flores_PRBR2016,Mazin_PRB_2015_SH3,Heil_PRB_2015_bonding} or
among alkali, alkaline-earth and rare-earth hydrides, which form \textit{sodalite} structures,
in which the $X$ atoms do not directly bind to hydrogen but provide a scaffold, 
that stabilizes a dense sponge-like hydrogen lattice \cite{Ma_PNAS_2012_CaH, Ma_PRL_2017_ReH, Heil_PRB_2019,Miao_RS_2021_ChemTemplate, Yi_PRM_2021_LaH10}.
Most HTSC binary hydrides are predicted to be stable above 150-200 GPa; 
a few are predicted to survive down to 70 GPa, where they are on the verge of a dynamical instability \cite{Duan_arXiv_2020_ReH, Heil_PRB_2019}; uranium hydrides are stable above 35 GPa, but do not exhibit HTSC \cite{Guigue_PRB_2020_UH7, Oganov_Science_2018_UH}. 

Having exhausted all possible combinations of binary hydrides, it is natural to extend the search for HTSC to multinary hydrides, where the addition of a third element other than hydrogen, enormously expands the phase space ~\cite{Kokail_PRM_2017_LiBH,  DiCataldo_PRB_2020_CaBH}.
A few works have already tried to exploit this additional flexibility to
to raise the $T_c$ of high-pressure hydrides well beyond room temperature.~\cite{Ma_PRL_2019_Li2MgH16,Oganov_arXiv_LaYH_2020}

%% UH_7 prediction and experiment: room pressure, predicted Tc 57-66 K. I. Kruglov, et al. SCIENCE ADVAN‏CES 4 (2018)
%% UH8 : same space group as LaBH8, same H Wyckoff, 
%% UH_7 and UH_8, experiment (down to 35 GPa). B. Guiguem et al. Phys. Rev. B 102, 014107 (2020)
%% ThH_7 ThH_10 70,80 K, and 150-170 GPa, D. Semenok, Materials Today 33, 2020
%

In this work, we will demonstrate a strategy to bring the stabilization pressure of high-\tc{} superconducting hydride phases close to ambient pressure in a ternary hydride. In short, it consists of identifying a suitable combination of elements with different sizes and electronegativity.  Our strategy is demonstrated by the prediction of a new ternary high-temperature superconductor, identified through a evolutionary search of the lanthanum-boron-hydrogen (La-B-H)
phase diagram \cite{USPEX_1, USPEX_2}. This phase, with LaBH$_8$ composition, exhibits a superconducting \tc{} of 126 K at 50 GPa. It is a remarkable example of a \textit{ternary sodalite hydride}, in which a cubic La-B scaffolding confines a highly-symmetric, dense hydrogen sublattice, and makes it stable down to moderate pressures.
Ternary sodalite hydride phases have been predicted only at pressures above 170 GPa \cite{Ma_PRL_2019_Li2MgH16, Liang_PRB_2019_CaYH},
while LaBH$_8$ appears on the ternary La-B-H hull at 110 GPa, and is dynamically stable down to 40 GPa.
%% Li2MgH12 473 K at 250 GPa

%% APPROX 170 WORDS
\begin{figure}[htpb]
	\centering
	\includegraphics[width=0.90\columnwidth]{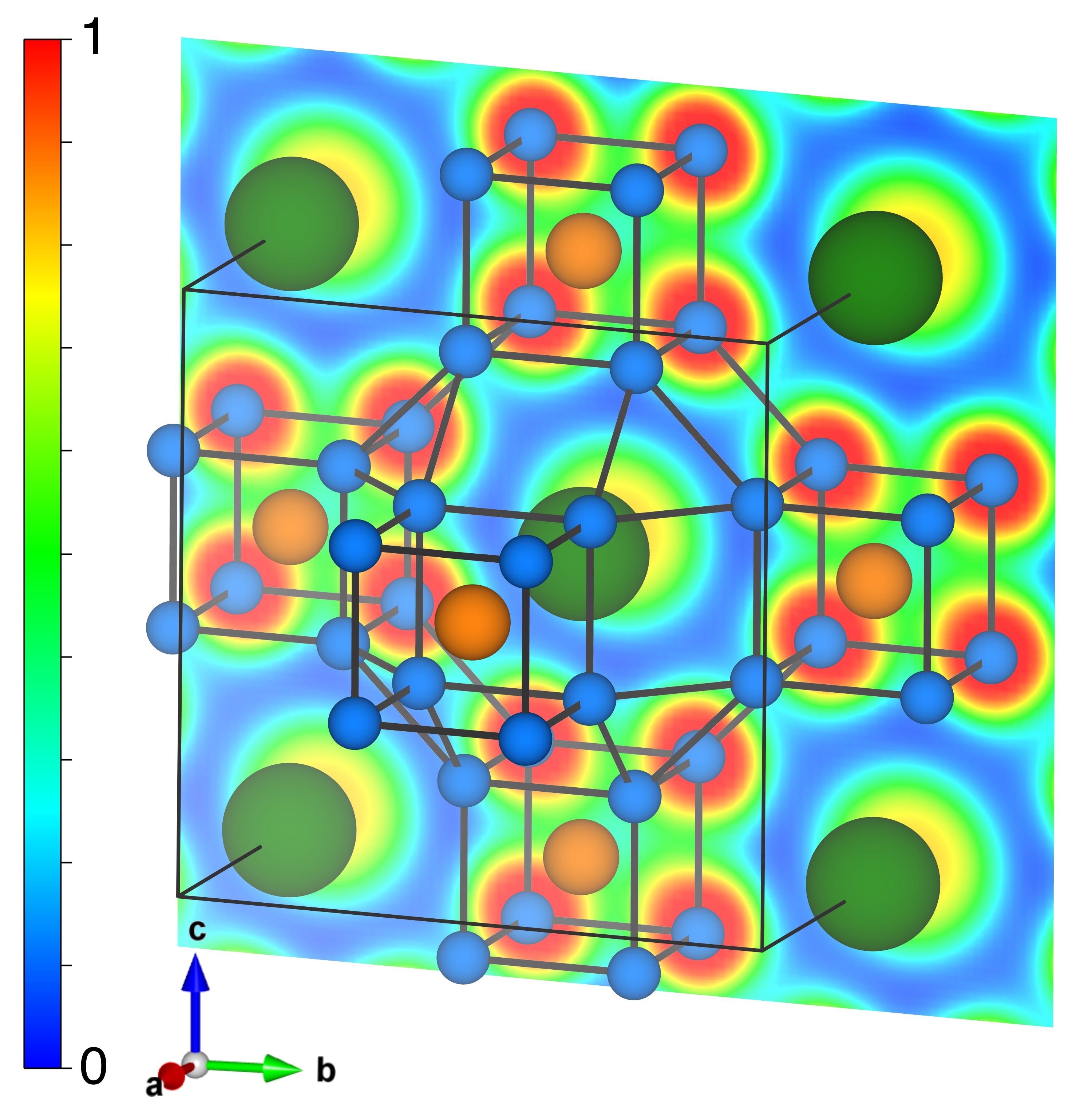}
	\caption{Crystal structure of the $Fm\bar{3}m$ phase of LaBH$_8$ (conventional unit cell). La, B, and H atoms are shown as green, orange, and blue spheres, respectively. The Electron Localization Function (ELF) is projected onto the $\bar{1}00$ plane.}
	\label{fig:labh8elf}
\end{figure}
%% STRUCTURAL PROPERTIES AND PHASE DIAGRAM
The new LaBH$_8$ phase was identified through a structural search at 100 GPa using variable-composition evolutionary crystal structure prediction, as implemented in USPEX \cite{USPEX_1, USPEX_2}, by sampling a total of over twelve thousand structures \footnote{In addition, we re-sampled particularly promising compositions. Further details are provided in the Supplemental Material}. The ternary La-B-H convex hull was then constructed, including also the zero-point lattice contribution to the total energy.
At 100 GPa four stable compositions lie on the convex hull: La(BH$_2$)$_3$, La(BH$_4$)$_3$, LaBH$_5$, and LaBH$_8$.
The first two exhibit crystal structures analogous to those observed in other metal borohydrides,
i.e.  molecular structures with BH$_{2}^{-}$ and BH$_{4}^{-}$ anions interspaced by La$^{3+}$ cations \cite{Jensen_ChemSocRev_2017_MBH, Wolverton_PRB_2010_CaBHx, DiCataldo_PRB_2020_CaBH, Jensen_IC_2020_LaB3H12, Jensen_PSS_2011_MBH}, and are insulating.
LaBH$_5$ and LaBH$_8$ are characterized by the same La-B rocksalt sublattice. In LaBH$_5$, boron and hydrogen form a BH$_{4}^{-}$ tetrahedral anion, and an additional H atom is trapped at the center of a La tetrahedron.
For LaBH$_8$ we predict a $Fm\bar{3}m$ sodalite-like structure.
 Both LaBH$_5$ and  LaBH$_8$ are metallic. Preliminary \tc{} calculations showed that the LaBH$_5$ structure exhibits a \tc{} of 53 K at 50 GPa, whereas LaBH$_{8}$ exhibits a \tc{} of 126 K at the same pressure \footnote{Additional information on the crystal structures can be found in the form of Crystallographic Information File in the Supplemental Material}.
These preliminary results led us to focus on the much more promising LaBH$_8$ $Fm\bar{3}m$ sodalite-like structure.

$Fm\bar{3}m$ LaBH$_8$ lies only 23 meV/atom above the hull at 100 GPa, becomes thermodynamically stable above 110 GPa, and is dynamically stable down to 40 GPa. This implies that this phase can be realistically synthesized by laser heating at 110 GPa, and quenched at low pressures down to a minimum of 40 GPa.
The crystal structure of $Fm\bar{3}m$-LaBH$_8$ is shown in Fig. \ref{fig:labh8elf}: La and B occupy $4b$ and $4a$ Wyckoff positions, respectively, while H atoms sit on $32f$ sites with $x = 0.145$. The hydrogen sites occupy the vertices of a rhombicuboctahedron centered around La atoms, and vertices of cubes around B atoms. Interestingly, a structure with an identical $M$-B sublattice was observed with neutron diffraction on $M$BH$_4$ ($M$ = K, Rb, Cs), which only differs by the $1/2$ occupancy of the $32f$ site by hydrogen \cite{Yvon_JAC_2004_structure_CsBH4}.

The Electron Localization Function (ELF) for LaBH$_{8}$, shown in Fig. \ref{fig:labh8elf} along the $\bar{1}00$ plane, has maxima around the atoms, and along the H-H bonds, but not between La and H or B and H, indicating that neither La nor B form bonds with H, but both act as spacers. The absence of a B-H covalent bond is quite unusual for a borohydride, and implies that this structure is not a covalent hydride, like SH$_{3}$.
Rather, it is reminiscent sodalite hydrides like LaH$_{10}$, where
a dense, metallic hydrogen sublattice is stabilized at pressures lower than
the pure hydrogen at metallization pressure, due to the chemical pressure exerted by the host atoms.
In a very simplified picture, one could see LaBH$_{8}$ as a chemically-precompressed version of LaH$_{10}$ (for a visual impression see Fig. S3 of the SM \cite{Note3}). In fact, the two structures share the same La-La sublattice, with almost identical lattice parameters at all pressures; LaH$_{10}$ is stable at the harmonic level only above 200 GPa; in LaBH$_{8}$, boron atoms fill the voids between the second-nearest La atoms and provide additional chemical pressure,
making the metallic hydrogen sublattice stable down to 40 GPa.
 The analogy of LaBH$_8$ with binary sodalite structures, which is confirmed by the analysis of the electronic structure and vibrational properties, is ultimately at the heart of the HTSC at low pressure. We also observe that at all pressures H-H interatomic distances are 13\% larger than sodalite LaH$_6$, and 20\% larger than LaH$_{10}$ \cite{Heil_PRB_2019} (see Fig. S5 of the SM  \footnote{The Supplemental Material is available at..}). In short, both the geometric and bonding properties indicate that this structure is a natural extension of the concept of \textit{sodalite} hydrides, to the case of a ternary hydride \cite{Ma_PNAS_2012_CaH, Ma_PRL_2017_ReH}.
%% LaH10 50 GPa: 5.5 A, 300 GPa: 4.8 A
%% LaBH8 50 GPa: 5.6 A, 300 GPa: 4.7 A

\begin{figure}[htpb]
	\centering
	\includegraphics[width=1.00\columnwidth]{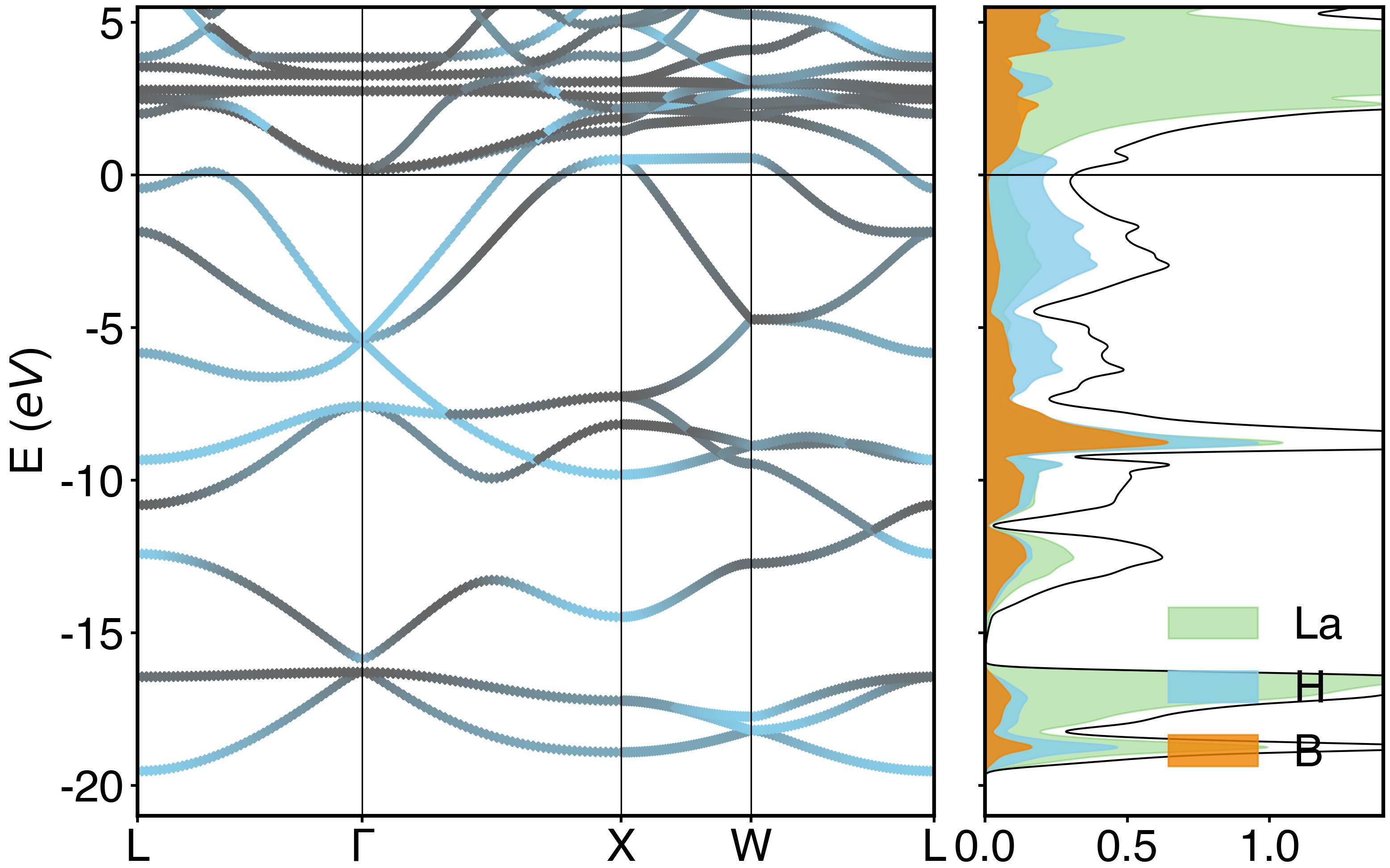}
	\caption{Left panel: electronic band structure of $Fm\bar{3}m$ LaBH$_8$ at 50 GPa, decorated with hydrogen character (blue) vs non-hydrogen character (gray). Right panel: atom-projected density of states in units of eV$^{-1}$spin$^{-1}$. Projection onto La, B, and H is shown in green, orange, and blue, respectively. The zero of the energy corresponds to the Fermi level.}
	\label{fig:labh8bands50gpa}
\end{figure}

%% ELECTRONIC STRUCTURE: BANDS; BADER CHARGE; DOS;
In Fig. \ref{fig:labh8bands50gpa} we show the electronic band structure, along with the atom-projected density of states, calculated at 50 GPa.
Here and in the following, we will focus on this pressure, which is sufficiently close to the moderate-pressure regime, but is about 10 GPa higher than the dynamical instability pressure, so that predictions are still not dramatically affected by anharmonic effects. 
A band structure formation analysis (see SM, Fig. S6 \cite{Note3})
reveals that the eight bands in the -15 to 2 eV range from the Fermi level derive from the eight quasi-free-electron-like  bands of the empty H$_{8}$ sublattice, which are only weakly perturbed by hybridization with the $2s-2p$ boron states,
and more strongly by hybridization with the three La semi-core bands from -20 to -15 eV.
A Bader charge analysis \cite{Henkelman_JCC_2007_Bader} predicts a net charge of +1.46 for La, +0.88 for B, and -0.29 for each H,
indicating that both boron and lanthanum donate  charge
to the hydrogen sublattice.

Hence, the band structure analysis confirms the bonding picture observed in real space: H forms a dense sublattice,
stabilized by the La-B scaffolding with which hydrogen forms only weak bonds. The absence of a covalent B-H bond here is crucial, and explains the free-electron-like behavior of the hydrogen-derived electronic states. In fact, as a result of this weak hybridization electronic bands at the Fermi level are highly-dispersive and have an almost purely (70\%) hydrogen character, exactly like binary sodalite hydrides \cite{Heil_PRB_2019}.

\begin{figure}[htpb]
	\centering
	\includegraphics[width=1.00\columnwidth]{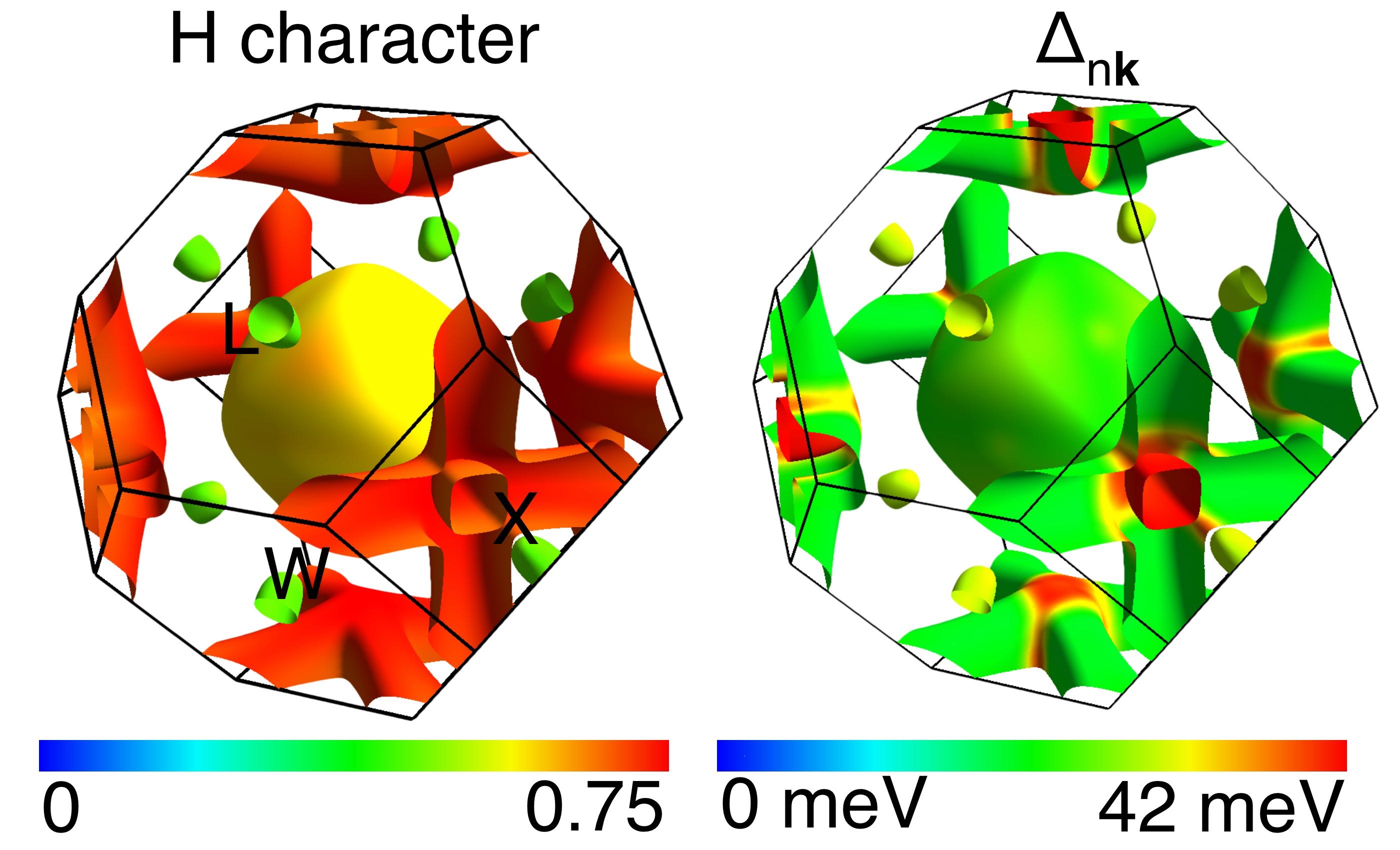}
	\caption{Fermi surface of LaBH$_8$ at 50 GPa. Left: decorated with hydrogen character, right: decorated with the value of the superconducting gap. The color scale goes from zero to the maximum value of the H character (0 to 0.75), and the gap (0 to 42 meV), respectively.}
	\label{fig:anisogaphcharacter}
\end{figure}

In the left panel of Fig.~\ref{fig:anisogaphcharacter} we show the Fermi surface decorated with H character.
The Fermi surface is characterized by three sheets i) a large electron-like sphere around the $\Gamma$ point, which has the greatest weight in the reciprocal space,
ii) a cross-shaped sheet and a small hole pocket around the X point, mostly of H character, and iii) a small hole pocket around the L point with mixed B and H character. Overall, the whole Fermi surface exhibits a strong hydrogen character,
i.e. the partial H contribution to the DOS is never less than 50 \%, on average around 70\%.

%% APPROX 114 WORDS

%% SUPERCONDUCTIVITY

\begin{figure}[htpb]
	\centering
	\includegraphics[width=1.00\columnwidth]{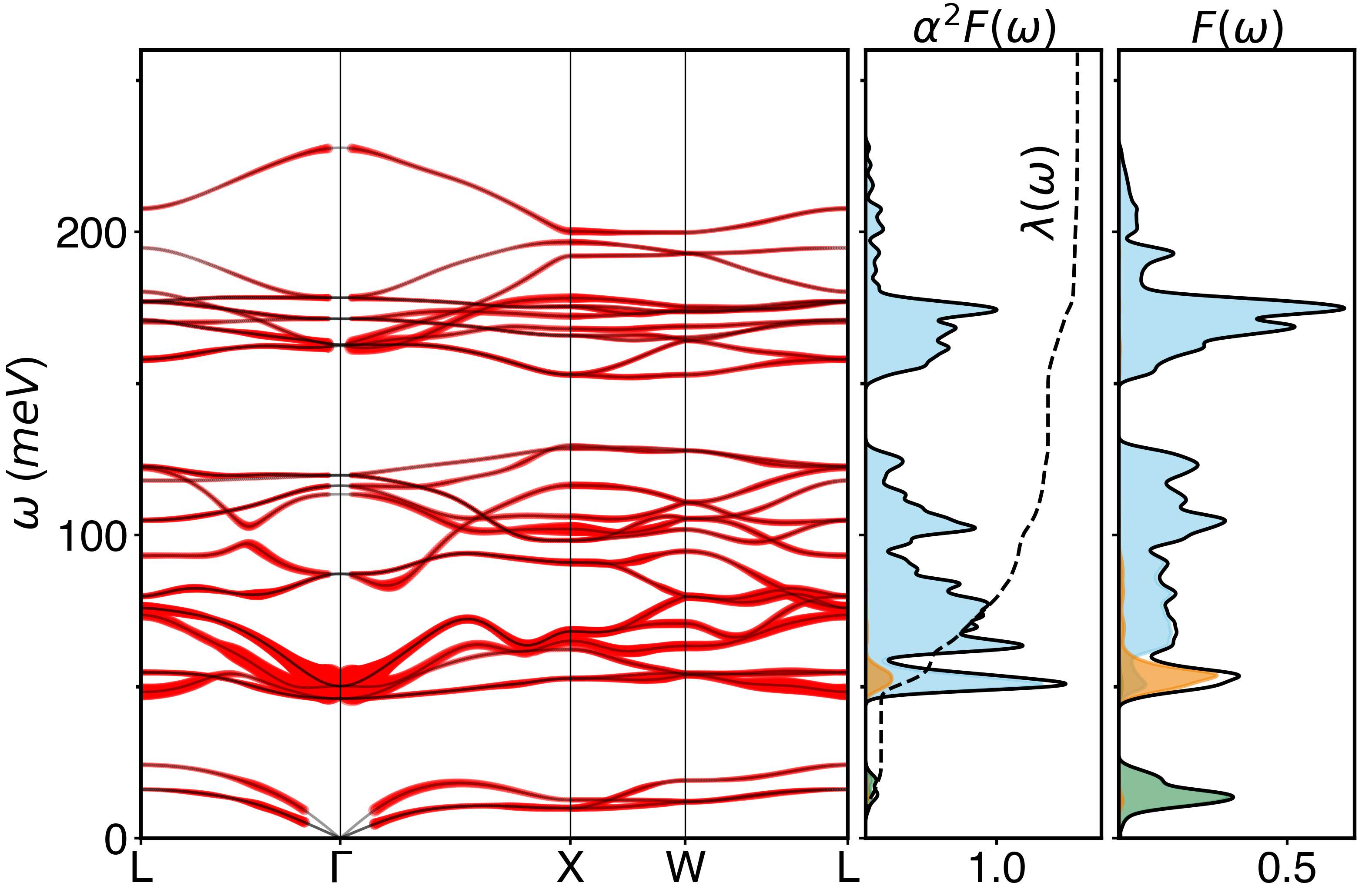}
	\caption{Left panel: phonon dispersions of LaBH$_8$ at 50 GPa (black thin lines), decorated with the \ep{} coupling (red thick lines). Center panel: atom-projected (colored filled lines) and total (black line) Eliashberg function, and its first inverse moment $\lambda(\omega)$ (dashed black line). Right panel: atom-projected (colored filled lines) and total (black line) phonon density of states. Projection onto La, B, and H is shown in green, orange, and blue, respectively.}
	\label{fig:labh8phbandsphwidths50gpa}
\end{figure}

In order to compute the superconducting properties of the $Fm\bar{3}m$-LaBH$_8$ phase,
we calculated the phonon dispersions and the electron-phonon coupling using linear response theory within the harmonic approximation,
using Wannier interpolation on very fine $\vec{k}$ and $\vec{q}$ grids, as implemented in the \verb*|EPW| code \cite{Baroni_RevModPhys_2001_DFPT, Savrasov_PRB_1996_lrt, Giustino_CPC_2016_EPW}.
In Fig. \ref{fig:labh8phbandsphwidths50gpa} we show the phonon dispersions decorated with the partial electron-phonon (\ep{}) coupling coefficients $\lambda_{\nu \vec{q}}$, together with the atom-projected Eliashberg function
$\alpha^2 F(\omega)$
and the phonon density of states. The  \ep{}  coupling is spread rather evenly on all optical branches, and is stronger
for modes which involve vibrations of the hydrogen sublattice, again in close analogy with other binary sodalite hydrides \cite{Mauri_Nature_2020_LaH, Heil_PRB_2019}.
The high peak in the Eliashberg function at 50 meV corresponds to a flat region of the dispersion around  the $\Gamma$ point, which
experiences a particularly strong \ep{} coupling. This phonon mode, named $T^{*}_{2g}$ in the following,  is characterized by a $T_{2g}$ symmetry  at the $\Gamma$ point and corresponds to a distortion of the tetrahedra formed by nearest-neighbours H atoms, and carries around 15$\%$ of the total \ep{} coupling (See SM Fig. S13 for more details \cite{Note3}).
In addition,  a triply degenerate branch with $T_{1u}$ symmetry at $\Gamma$,  accidentally quasi-degenerate with the $T_{2g}^{*}$ mode at 50 GPa, is also notable.
This branch, in fact,  corresponds to a rattling mode of the boron atoms inside the cubic hydrogen cages surrounding it, and is mostly dispersionless throughout the Brillouin zone, coherently with the description of boron as passively pressurizing the metallic hydrogen sublattice, without bonding to it.

Integrating the Eliashberg function we obtain the two moments: \cite{McMillan_PR_1968, Allen_PRB_1975_McMillan}
$\lambda =  2\int{\alpha^2F(\omega)\omega^{-1} d\omega} = 1.54$
  and $\omega_{\text{log}} = \exp \left[ 2\lambda^{-1} \int d\omega  \alpha^2F( \omega)\omega^{-1} \log( \omega) \right] 71$ meV.
Table  \ref{tab:superproperties}  reports the critical temperature obtained by a direct solution of ab-initio Migdal-Eliashberg equations, as implemented in the \verb*|EPW| code \cite{Giustino_CPC_2016_EPW}.
Coulomb effects are included via the Morel-Anderson pseudopotential $\displaystyle \mu^{*} = \mu/[1 + \mu\log(\omega_{\text{el}}/\omega_{\text{ph}})]$ \cite{Morel_PhysRev_1962_mustar}, with $\omega_{\text{el}}$ and $\omega_{\text{ph}}$ being characteristic energies for electrons (band-width of the Fermi surface electrons) and phonons (highest phonon energy), respectively. The double Fermi-surface average of the screened Coulomb interaction $\mu$ was evaluated within the random phase approximation \cite{Lee_PRB_1995_mustar, Giustino_PRB_2010_GWsternheimer, Lambert_PRB_2013_GWsternheimer, Heil_PRL_2017_NbS2}. We find a value of $\mu^{*}$=0.09 at pressures of 50 and 100 GPa, close to the standard values (0.10-0.14) assumed for most conventional superconductors~\cite{Heil_PRB_2019}. This rules out possible anomalous effects of Coulomb repulsion which were suggested for yttrium sodalite hydride \cite{Oganov_arXiv_2019_YH6}.

The \tc{} obtained from the fully anisotropic solution (\tc{}=126 K) is extremely close to the isotropic one (\tc{}=122 K). The anisotropy of the superconducting gap is in fact very limited \footnote{We also checked the dependence of $\mu^{*}$ on Tc and found that a variation of 0.01 in $\mu^{*}$ changes Tc only by 2-3 K (See SM Fig. S9 for more details) \cite{Note3}.}
The distribution of the calculated superconducting gap on the Fermi surface is shown in the right panel of Fig. \ref{fig:anisogaphcharacter}.
Indeed, with the exception of a \textit{hotspot} around the X point, which has a negligible
weigth in reciprocal space, the gap is rather uniform, with a deviation of $\pm$ 15\% around its mean value $\Delta_{\text{avg}}$ = 26 meV. The mean value differs from the isotropic average $\Delta_{\text{iso}} =$ 23.5 meV, as the isotropic average is affected by the fact that large values of the gap have a small weight in reciprocal space. The calculated BCS parameter $\displaystyle 2 \Delta_{\text{iso}}(0)/T_{c}$ is 4.3, confirming the strong-coupling nature of superconductivity in LaBH$_{8}$. 

%% APPROX 118 WORDS

%% 32.5 WORDS (13 + 6.5 per line)
\begin{table}[htpb]
\begin{tabular}{cccccccc}
\hline %% OFFICIAL PRL TABLE FORMAT
\hline
\addlinespace[0.1cm]
P        &      $N(E_{F})$      &     $\lambda$     &    $\omega_{\text{log}}$			 &   T$_c^{\text{AME}}$  & T$_c^{\text{IME}}$ & $\Delta_{\text{iso}}$ & $\Delta$H\\
(GPa) 	&	$(eV^{-1})$&							&		 (meV)					&				(K)					&				 (K)		& (meV) &		(meV/at)	\\
\hline
50                &          0.62        &           1.54        &             71            &         126    &    122   &	23.5 	& 125	 \\
75                &           0.60       &            1.06        &           91            &          101     &     96   &	16.8	  &  71	 \\
100               &          0.56    &           0.64        &            88            &          42    &     32    &	5.6	 &	23 \\
\hline
\hline
\end{tabular}
\caption{Summary of the main superconducting properties of LaBH$_{8}$ at 50 and 100 GPa. The DOS at the Fermi level $N(E_{F}$ in the second column is in units of $eV^{-1}spin^{-1}$. \textit{AME} and \textit{IME} correspond to solutions of the anisotropic and isotropic Migdal-Eliashberg equations, respectively. The \tc{} is calculated with $\mu^{*}=0.09$. $\Delta_{\text{iso}}$ represents the isotropic average of the superconducting gap. The last column describes the enthalpy per atom (including zero-point energy) above the convex hull for the LaBH$_{8}$ $Fm\bar{3}m$ phase.}
\label{tab:superproperties}
\end{table}

%% APPROX 111 WORDS

Having established that the superconducting properties of LaBH$_{8}$ at 50 GPa are extremely
promising, we further studied their behavior as a function of pressure,
computing the electron-phonon spectra at 75 and 100 GPa, and solving the
corresponding Eliashberg equations. The main results are summarized in Tab. \ref{tab:superproperties}, and more details are provided in SM Fig. S8 \cite{Note3}.
The main effect of an increase in pressure is a rather uniform shift of all phonon frequencies to higher values,
which causes a decrease of $\lambda$ and an increase of $\omega_{log}$. 
The $T_{2g}^{*}$ mode around $\Gamma$, which at 50 GPa has a frequency of 50 meV, responds
more strongly to pressure than the rest of the spectrum, causing a small, counterintuitive decrease of $\omega_{\text{log}}$
between 75 and 100 GPa. The same mode drives the system towards a dynamical instability when pressure is decreased below 50 GPa.

%% \footnote{By analyzing the wave vector and mode-resolved \ep{} coupling $\lambda_{\nu \vec{q}}$ as well as the wave vector resolved nesting function $\zeta_0$, as detailed in SM Fig. S9 \cite{Note3}, we find that Fermi surface nesting alone cannot account for the observed phonon softening, pointing to a strong \ep{} coupling for certain $\nu$ and $\vec{q}$.}. 
%
At the harmonic level, the instability occurs at 35 GPa. Vibrations involving hydrogen and, in general, light elements, exhibit strong anharmonic and quantum effects \cite{Mauri_Nature_2016_SH3, Mauri_Nature_2020_LaH},
which may severely affect the dynamical stability and/or superconducting properties of hydrides. In LaBH$_{8}$, the soft $T^{*}_{2g}$ mode is also the only strongly anharmonic one. Hence, in order to estimate the importance of anharmonic effects in $Fm\bar{3}m$-LaBH$_8$, 
We recomputed the frequency of the $T_{2g}^{*}$ mode,
solving the Schr\"{o}dinger equation numerically as a function of pressure, as described in Ref. \cite{Heil_PRL_2017_NbS2}.
We estimated that the difference between the harmonic and anharmonic frequencies is
almost constant with pressure, and equal to $\sim 10$ meV. This causes a $\sim 5$ GPa shift of the stability pressure to 40 GPa (See Figs. S11 and S12 of the SM for more details \cite{Note3}), and a negligible effect on the critical temperature (See Fig. S13 of the SM \cite{Note3}).
The stabilization pressure of LaBH$_8$ represents a new minimum at which a high-\tc{} superhydride is predicted to be stable, beating the previous record of 70 GPa in YbH$_6$ \cite{Duan_arXiv_2020_ReH}. We believe that the main reason behind the low stabilization pressure of LaBH$_8$ is \textit{chemical pressure}.

%% CONCLUSIONS
In conclusion, using a evolutionary crystal structure prediction and ab-initio Migdal Eliashberg theory we predicted a new ternary hydride phase with LaBH$_{8}$ stoichiometry and $Fm\bar{3}m$ space group, which is a conventional HTSC at moderate pressures, with a \tc{} of 126 K at 50 GPa. According to our estimate, this structure could be synthesized by means of laser heating at a pressure around 100 GPa, and remains dynamically stable down to 40 GPa, where a single zone-center phonon mode drives a structural instability.

LaBH$_{8}$ is the first conventional superconductor with \tc{} above liquid nitrogen boiling point that can be stabilized down to 50 GPa. Its exceptional superconducting properties can be understood as deriving from a metallic hydrogen lattice, which is stabilized at low pressures by a boron and lanthanum scaffolding. The combination of two elements with different atomic sizes turns out to be a very effective way to boost chemical pressure on the interstitial hydrogen sublattice. In general, our results demonstrate an effective new strategy to lower the stabilization pressure of binary hydrides. It is likely that the $XY$H$_{8}$ $Fm\bar{3}m$ structure may be tuned to attain even better performances, through a careful choice of the $X$,$Y$ elements.
 The possibility of stabilizing a superhydride to this pressure represents a giant leap towards hydride-based superconductivity at room pressure. 

\section{Acknowledgments}
This work was supported by the Austrian Science Fund (FWF) Projects No. P 30269-N36 (Superhydra), the dCluster of the Graz University of Technology and the VSC3 of the Vienna University of Technology. L.B. acknowledges support from Fondo Ateneo- Sapienza 2017,2018 and 2019. C. H. acknowledges support from the Austrian Science Fund (FWF) Project No. P 32144-N36 and the VSC4 of the Vienna University of Technology. The authors would like to thank Antonio Sanna for the useful suggestions and for kindly sharing the code to solve the isotropic Migdal-Eliashberg equations.

%% MAIN TEXT: MAX 3750 WORDS
%% ABSTRACT: 165 (NOT COUNTED IN THE TOTAL)
%% CURRENT COUNT: 
%% MAIN: 2753
%% FIGURES: approx. 513
%% TABLES: approx. 32.5
%% EQUATIONS: 32
%% TOTAL: 3331
\bibliographystyle{apsrev4-2.bst}

\begin{thebibliography}{51}%
	\makeatletter
	\providecommand \@ifxundefined [1]{%
		\@ifx{#1\undefined}
	}%
	\providecommand \@ifnum [1]{%
		\ifnum #1\expandafter \@firstoftwo
		\else \expandafter \@secondoftwo
		\fi
	}%
	\providecommand \@ifx [1]{%
		\ifx #1\expandafter \@firstoftwo
		\else \expandafter \@secondoftwo
		\fi
	}%
	\providecommand \natexlab [1]{#1}%
	\providecommand \enquote  [1]{``#1''}%
	\providecommand \bibnamefont  [1]{#1}%
	\providecommand \bibfnamefont [1]{#1}%
	\providecommand \citenamefont [1]{#1}%
	\providecommand \href@noop [0]{\@secondoftwo}%
	\providecommand \href [0]{\begingroup \@sanitize@url \@href}%
	\providecommand \@href[1]{\@@startlink{#1}\@@href}%
	\providecommand \@@href[1]{\endgroup#1\@@endlink}%
	\providecommand \@sanitize@url [0]{\catcode `\\12\catcode `\$12\catcode
		`\&12\catcode `\#12\catcode `\^12\catcode `\_12\catcode `\%12\relax}%
	\providecommand \@@startlink[1]{}%
	\providecommand \@@endlink[0]{}%
	\providecommand \url  [0]{\begingroup\@sanitize@url \@url }%
	\providecommand \@url [1]{\endgroup\@href {#1}{\urlprefix }}%
	\providecommand \urlprefix  [0]{URL }%
	\providecommand \Eprint [0]{\href }%
	\providecommand \doibase [0]{https://doi.org/}%
	\providecommand \selectlanguage [0]{\@gobble}%
	\providecommand \bibinfo  [0]{\@secondoftwo}%
	\providecommand \bibfield  [0]{\@secondoftwo}%
	\providecommand \translation [1]{[#1]}%
	\providecommand \BibitemOpen [0]{}%
	\providecommand \bibitemStop [0]{}%
	\providecommand \bibitemNoStop [0]{.\EOS\space}%
	\providecommand \EOS [0]{\spacefactor3000\relax}%
	\providecommand \BibitemShut  [1]{\csname bibitem#1\endcsname}%
	\let\auto@bib@innerbib\@empty
	%</preamble>
	\bibitem [{\citenamefont {Einaga}\ \emph {et~al.}(2016)\citenamefont {Einaga},
		\citenamefont {Sakata}, \citenamefont {Ishikawa}, \citenamefont {Shimizu},
		\citenamefont {Eremets}, \citenamefont {Drodzov}, \citenamefont {Troyan},
		\citenamefont {Hirao},\ and\ \citenamefont
		{Ohishi}}]{Eremets_NatPhys_2016_SH3}%
	\BibitemOpen
	\bibfield  {author} {\bibinfo {author} {\bibfnamefont {M.}~\bibnamefont
			{Einaga}}, \bibinfo {author} {\bibfnamefont {M.}~\bibnamefont {Sakata}},
		\bibinfo {author} {\bibfnamefont {T.}~\bibnamefont {Ishikawa}}, \bibinfo
		{author} {\bibfnamefont {K.}~\bibnamefont {Shimizu}}, \bibinfo {author}
		{\bibfnamefont {M.}~\bibnamefont {Eremets}}, \bibinfo {author} {\bibfnamefont
			{A.~P.}\ \bibnamefont {Drodzov}}, \bibinfo {author} {\bibfnamefont {I.~A.}\
			\bibnamefont {Troyan}}, \bibinfo {author} {\bibfnamefont {N.}~\bibnamefont
			{Hirao}},\ and\ \bibinfo {author} {\bibfnamefont {Y.}~\bibnamefont
			{Ohishi}},\ }\href@noop {} {\bibfield  {journal} {\bibinfo  {journal} {Nature
				Physics}\ }\textbf {\bibinfo {volume} {12}},\ \bibinfo {pages} {835}
		(\bibinfo {year} {2016})}\BibitemShut {NoStop}%
	\bibitem [{\citenamefont {Drodzov}\ \emph
		{et~al.}(2015{\natexlab{a}})\citenamefont {Drodzov}, \citenamefont {Eremets},
		\citenamefont {Troyan}, \citenamefont {Ksenofontov},\ and\ \citenamefont
		{Shylin}}]{Eremets_Nature_2015_SH3}%
	\BibitemOpen
	\bibfield  {author} {\bibinfo {author} {\bibfnamefont {A.~P.}\ \bibnamefont
			{Drodzov}}, \bibinfo {author} {\bibfnamefont {M.~I.}\ \bibnamefont
			{Eremets}}, \bibinfo {author} {\bibfnamefont {I.~A.}\ \bibnamefont {Troyan}},
		\bibinfo {author} {\bibfnamefont {V.}~\bibnamefont {Ksenofontov}},\ and\
		\bibinfo {author} {\bibfnamefont {S.~I.}\ \bibnamefont {Shylin}},\
	}\href@noop {} {\bibfield  {journal} {\bibinfo  {journal} {Nature}\ }\textbf
		{\bibinfo {volume} {525}},\ \bibinfo {pages} {73} (\bibinfo {year}
		{2015}{\natexlab{a}})}\BibitemShut {NoStop}%
	\bibitem [{\citenamefont {Duan}\ \emph {et~al.}(2014)\citenamefont {Duan},
		\citenamefont {Liu}, \citenamefont {Tian}, \citenamefont {Li}, \citenamefont
		{Huang}, \citenamefont {Zhao}, \citenamefont {Yu}, \citenamefont {Liu},
		\citenamefont {Tian},\ and\ \citenamefont {Cui}}]{Duan_SciRep_2014_SH}%
	\BibitemOpen
	\bibfield  {author} {\bibinfo {author} {\bibfnamefont {D.}~\bibnamefont
			{Duan}}, \bibinfo {author} {\bibfnamefont {Y.}~\bibnamefont {Liu}}, \bibinfo
		{author} {\bibfnamefont {F.}~\bibnamefont {Tian}}, \bibinfo {author}
		{\bibfnamefont {D.}~\bibnamefont {Li}}, \bibinfo {author} {\bibfnamefont
			{X.}~\bibnamefont {Huang}}, \bibinfo {author} {\bibfnamefont
			{Z.}~\bibnamefont {Zhao}}, \bibinfo {author} {\bibfnamefont {H.}~\bibnamefont
			{Yu}}, \bibinfo {author} {\bibfnamefont {B.}~\bibnamefont {Liu}}, \bibinfo
		{author} {\bibfnamefont {W.}~\bibnamefont {Tian}},\ and\ \bibinfo {author}
		{\bibfnamefont {T.}~\bibnamefont {Cui}},\ }\href@noop {} {\bibfield
		{journal} {\bibinfo  {journal} {Scientific Reports}\ }\textbf {\bibinfo
			{volume} {4}},\ \bibinfo {pages} {6968} (\bibinfo {year} {2014})}\BibitemShut
	{NoStop}%
	\bibitem [{\citenamefont {Kong}\ \emph {et~al.}(2019)\citenamefont {Kong},
		\citenamefont {Minkov}, \citenamefont {Kuzonikov}, \citenamefont {Besedin},
		\citenamefont {Drodzov}, \citenamefont {Mozaffari}, \citenamefont {Balicas},
		\citenamefont {Balakirev}, \citenamefont {Prakapenka}, \citenamefont
		{Greenberg}, \citenamefont {Knyazev},\ and\ \citenamefont
		{Eremets}}]{Eremets_arXiv_2019_YH6}%
	\BibitemOpen
	\bibfield  {author} {\bibinfo {author} {\bibfnamefont {P.~P.}\ \bibnamefont
			{Kong}}, \bibinfo {author} {\bibfnamefont {V.~S.}\ \bibnamefont {Minkov}},
		\bibinfo {author} {\bibfnamefont {M.~A.}\ \bibnamefont {Kuzonikov}}, \bibinfo
		{author} {\bibfnamefont {S.~P.}\ \bibnamefont {Besedin}}, \bibinfo {author}
		{\bibfnamefont {A.~P.}\ \bibnamefont {Drodzov}}, \bibinfo {author}
		{\bibfnamefont {S.}~\bibnamefont {Mozaffari}}, \bibinfo {author}
		{\bibfnamefont {L.}~\bibnamefont {Balicas}}, \bibinfo {author} {\bibfnamefont
			{F.~F.}\ \bibnamefont {Balakirev}}, \bibinfo {author} {\bibfnamefont {V.~B.}\
			\bibnamefont {Prakapenka}}, \bibinfo {author} {\bibfnamefont
			{E.}~\bibnamefont {Greenberg}}, \bibinfo {author} {\bibfnamefont {D.~A.}\
			\bibnamefont {Knyazev}},\ and\ \bibinfo {author} {\bibfnamefont {M.~I.}\
			\bibnamefont {Eremets}},\ }\href@noop {} {\bibfield  {journal} {\bibinfo
			{journal} {arXiv:1909.10482}\ } (\bibinfo {year} {2019})}\BibitemShut
	{NoStop}%
	\bibitem [{\citenamefont {Troyan}\ \emph {et~al.}(2019)\citenamefont {Troyan},
		\citenamefont {Semenok}, \citenamefont {Kvashin}, \citenamefont {Sadakov},
		\citenamefont {Sobolevskiy}, \citenamefont {Pudalov}, \citenamefont
		{Ivanova}, \citenamefont {Prakapenka}, \citenamefont {Greenberg},
		\citenamefont {Gavriliuk}, \citenamefont {Struzhkin}, \citenamefont
		{Bergara}, \citenamefont {Errea}, \citenamefont {Bianco}, \citenamefont
		{Calandra}, \citenamefont {Mauri}, \citenamefont {Monacelli}, \citenamefont
		{Akashi},\ and\ \citenamefont {Oganov}}]{Oganov_arXiv_2019_YH6}%
	\BibitemOpen
	\bibfield  {author} {\bibinfo {author} {\bibfnamefont {I.~A.}\ \bibnamefont
			{Troyan}}, \bibinfo {author} {\bibfnamefont {D.~V.}\ \bibnamefont {Semenok}},
		\bibinfo {author} {\bibfnamefont {A.~G.}\ \bibnamefont {Kvashin}}, \bibinfo
		{author} {\bibfnamefont {A.~V.}\ \bibnamefont {Sadakov}}, \bibinfo {author}
		{\bibfnamefont {O.~A.}\ \bibnamefont {Sobolevskiy}}, \bibinfo {author}
		{\bibfnamefont {V.~M.}\ \bibnamefont {Pudalov}}, \bibinfo {author}
		{\bibfnamefont {A.~G.}\ \bibnamefont {Ivanova}}, \bibinfo {author}
		{\bibfnamefont {V.~B.}\ \bibnamefont {Prakapenka}}, \bibinfo {author}
		{\bibfnamefont {E.}~\bibnamefont {Greenberg}}, \bibinfo {author}
		{\bibfnamefont {A.~G.}\ \bibnamefont {Gavriliuk}}, \bibinfo {author}
		{\bibfnamefont {V.~V.}\ \bibnamefont {Struzhkin}}, \bibinfo {author}
		{\bibfnamefont {A.}~\bibnamefont {Bergara}}, \bibinfo {author} {\bibfnamefont
			{I.}~\bibnamefont {Errea}}, \bibinfo {author} {\bibfnamefont
			{R.}~\bibnamefont {Bianco}}, \bibinfo {author} {\bibfnamefont
			{M.}~\bibnamefont {Calandra}}, \bibinfo {author} {\bibfnamefont
			{F.}~\bibnamefont {Mauri}}, \bibinfo {author} {\bibfnamefont
			{L.}~\bibnamefont {Monacelli}}, \bibinfo {author} {\bibfnamefont
			{R.}~\bibnamefont {Akashi}},\ and\ \bibinfo {author} {\bibfnamefont {A.~R.}\
			\bibnamefont {Oganov}},\ }\href@noop {} {\bibfield  {journal} {\bibinfo
			{journal} {arXiv:1908.01534}\ } (\bibinfo {year} {2019})}\BibitemShut
	{NoStop}%
	\bibitem [{\citenamefont {Kruglov}\ \emph {et~al.}(2018)\citenamefont
		{Kruglov}, \citenamefont {Kvashin}, \citenamefont {Goncharov}, \citenamefont
		{Oganov}, \citenamefont {Lobanov}, \citenamefont {Holtgrewe}, \citenamefont
		{Jiang}, \citenamefont {Prakapenka}, \citenamefont {Greenberg},\ and\
		\citenamefont {Yanilkin}}]{Oganov_Science_2018_UH}%
	\BibitemOpen
	\bibfield  {author} {\bibinfo {author} {\bibfnamefont {I.~A.}\ \bibnamefont
			{Kruglov}}, \bibinfo {author} {\bibfnamefont {A.~G.}\ \bibnamefont
			{Kvashin}}, \bibinfo {author} {\bibfnamefont {A.~F.}\ \bibnamefont
			{Goncharov}}, \bibinfo {author} {\bibfnamefont {A.~R.}\ \bibnamefont
			{Oganov}}, \bibinfo {author} {\bibfnamefont {S.~S.}\ \bibnamefont {Lobanov}},
		\bibinfo {author} {\bibfnamefont {N.}~\bibnamefont {Holtgrewe}}, \bibinfo
		{author} {\bibfnamefont {S.}~\bibnamefont {Jiang}}, \bibinfo {author}
		{\bibfnamefont {V.~B.}\ \bibnamefont {Prakapenka}}, \bibinfo {author}
		{\bibfnamefont {E.}~\bibnamefont {Greenberg}},\ and\ \bibinfo {author}
		{\bibfnamefont {A.~V.}\ \bibnamefont {Yanilkin}},\ }\href@noop {} {\bibfield
		{journal} {\bibinfo  {journal} {Science Advances}\ }\textbf {\bibinfo
			{volume} {4}} (\bibinfo {year} {2018})}\BibitemShut {NoStop}%
	\bibitem [{\citenamefont {Semenok}\ \emph
		{et~al.}(2020{\natexlab{a}})\citenamefont {Semenok}, \citenamefont {Kvashin},
		\citenamefont {Ivanova}, \citenamefont {Svitlyk}, \citenamefont {Fominski},
		\citenamefont {Sadakov}, \citenamefont {Sobolevskiy}, \citenamefont
		{Pudalov}, \citenamefont {Troyan},\ and\ \citenamefont
		{Oganov}}]{Oganov_MatToday_2020_ThH}%
	\BibitemOpen
	\bibfield  {author} {\bibinfo {author} {\bibfnamefont {D.~V.}\ \bibnamefont
			{Semenok}}, \bibinfo {author} {\bibfnamefont {A.~G.}\ \bibnamefont
			{Kvashin}}, \bibinfo {author} {\bibfnamefont {A.~G.}\ \bibnamefont
			{Ivanova}}, \bibinfo {author} {\bibfnamefont {V.}~\bibnamefont {Svitlyk}},
		\bibinfo {author} {\bibfnamefont {V.~Y.}\ \bibnamefont {Fominski}}, \bibinfo
		{author} {\bibfnamefont {A.~V.}\ \bibnamefont {Sadakov}}, \bibinfo {author}
		{\bibfnamefont {O.~A.}\ \bibnamefont {Sobolevskiy}}, \bibinfo {author}
		{\bibfnamefont {V.~M.}\ \bibnamefont {Pudalov}}, \bibinfo {author}
		{\bibfnamefont {I.~A.}\ \bibnamefont {Troyan}},\ and\ \bibinfo {author}
		{\bibfnamefont {A.~R.}\ \bibnamefont {Oganov}},\ }\href@noop {} {\bibfield
		{journal} {\bibinfo  {journal} {Materials Today}\ }\textbf {\bibinfo {volume}
			{33}},\ \bibinfo {pages} {36} (\bibinfo {year}
		{2020}{\natexlab{a}})}\BibitemShut {NoStop}%
	\bibitem [{\citenamefont {Drodzov}\ \emph
		{et~al.}(2015{\natexlab{b}})\citenamefont {Drodzov}, \citenamefont
		{Eremets},\ and\ \citenamefont {Troyan}}]{Eremets_arXiv_PH3_2015}%
	\BibitemOpen
	\bibfield  {author} {\bibinfo {author} {\bibfnamefont {A.~P.}\ \bibnamefont
			{Drodzov}}, \bibinfo {author} {\bibfnamefont {M.~I.}\ \bibnamefont
			{Eremets}},\ and\ \bibinfo {author} {\bibfnamefont {I.~A.}\ \bibnamefont
			{Troyan}},\ }\href@noop {} {\bibfield  {journal} {\bibinfo  {journal}
			{arXiv:1508.06224}\ } (\bibinfo {year} {2015}{\natexlab{b}})}\BibitemShut
	{NoStop}%
	\bibitem [{\citenamefont {Drodzov}\ \emph {et~al.}(2019)\citenamefont
		{Drodzov}, \citenamefont {Kong}, \citenamefont {Besedin}, \citenamefont
		{Kuzonikov}, \citenamefont {Mozaffari}, \citenamefont {Balicas},
		\citenamefont {Balakirev}, \citenamefont {Graf}, \citenamefont {Prakapenka},
		\citenamefont {Greenberg}, \citenamefont {Knyazev}, \citenamefont {Tkacz},\
		and\ \citenamefont {Eremets}}]{Eremets_Nature_2019_LaH}%
	\BibitemOpen
	\bibfield  {author} {\bibinfo {author} {\bibfnamefont {A.~P.}\ \bibnamefont
			{Drodzov}}, \bibinfo {author} {\bibfnamefont {P.~P.}\ \bibnamefont {Kong}},
		\bibinfo {author} {\bibfnamefont {S.~P.}\ \bibnamefont {Besedin}}, \bibinfo
		{author} {\bibfnamefont {M.~A.}\ \bibnamefont {Kuzonikov}}, \bibinfo {author}
		{\bibfnamefont {S.}~\bibnamefont {Mozaffari}}, \bibinfo {author}
		{\bibfnamefont {L.}~\bibnamefont {Balicas}}, \bibinfo {author} {\bibfnamefont
			{F.~F.}\ \bibnamefont {Balakirev}}, \bibinfo {author} {\bibfnamefont {D.~E.}\
			\bibnamefont {Graf}}, \bibinfo {author} {\bibfnamefont {V.~B.}\ \bibnamefont
			{Prakapenka}}, \bibinfo {author} {\bibfnamefont {E.}~\bibnamefont
			{Greenberg}}, \bibinfo {author} {\bibfnamefont {D.~A.}\ \bibnamefont
			{Knyazev}}, \bibinfo {author} {\bibfnamefont {M.}~\bibnamefont {Tkacz}},\
		and\ \bibinfo {author} {\bibfnamefont {M.~I.}\ \bibnamefont {Eremets}},\
	}\href@noop {} {\bibfield  {journal} {\bibinfo  {journal} {Nature}\ }\textbf
		{\bibinfo {volume} {569}},\ \bibinfo {pages} {528} (\bibinfo {year}
		{2019})}\BibitemShut {NoStop}%
	\bibitem [{\citenamefont {Somayazulu}\ \emph {et~al.}(2019)\citenamefont
		{Somayazulu}, \citenamefont {Ahart}, \citenamefont {Mishra}, \citenamefont
		{Geballe}, \citenamefont {Baldini}, \citenamefont {Meng}, \citenamefont
		{Struzhkin},\ and\ \citenamefont {Hemley}}]{Hemley_PRL_2019_LaH}%
	\BibitemOpen
	\bibfield  {author} {\bibinfo {author} {\bibfnamefont {M.}~\bibnamefont
			{Somayazulu}}, \bibinfo {author} {\bibfnamefont {M.}~\bibnamefont {Ahart}},
		\bibinfo {author} {\bibfnamefont {A.~K.}\ \bibnamefont {Mishra}}, \bibinfo
		{author} {\bibfnamefont {Z.~M.}\ \bibnamefont {Geballe}}, \bibinfo {author}
		{\bibfnamefont {M.}~\bibnamefont {Baldini}}, \bibinfo {author} {\bibfnamefont
			{Y.}~\bibnamefont {Meng}}, \bibinfo {author} {\bibfnamefont {V.~V.}\
			\bibnamefont {Struzhkin}},\ and\ \bibinfo {author} {\bibfnamefont {R.~J.}\
			\bibnamefont {Hemley}},\ }\href@noop {} {\bibfield  {journal} {\bibinfo
			{journal} {Phys. Rev. Lett.}\ }\textbf {\bibinfo {volume} {122}},\ \bibinfo
		{pages} {027001} (\bibinfo {year} {2019})}\BibitemShut {NoStop}%
	\bibitem [{\citenamefont {Flores-Livas}\ \emph {et~al.}(2020)\citenamefont
		{Flores-Livas}, \citenamefont {Boeri}, \citenamefont {Sanna}, \citenamefont
		{Profeta}, \citenamefont {Arita},\ and\ \citenamefont
		{Eremets}}]{Boeri_PhysRep_2020_review}%
	\BibitemOpen
	\bibfield  {author} {\bibinfo {author} {\bibfnamefont {J.~A.}\ \bibnamefont
			{Flores-Livas}}, \bibinfo {author} {\bibfnamefont {L.}~\bibnamefont {Boeri}},
		\bibinfo {author} {\bibfnamefont {A.}~\bibnamefont {Sanna}}, \bibinfo
		{author} {\bibfnamefont {G.}~\bibnamefont {Profeta}}, \bibinfo {author}
		{\bibfnamefont {R.}~\bibnamefont {Arita}},\ and\ \bibinfo {author}
		{\bibfnamefont {M.}~\bibnamefont {Eremets}},\ }\href@noop {} {\bibfield
		{journal} {\bibinfo  {journal} {Physics Reports}\ }\textbf {\bibinfo {volume}
			{856}},\ \bibinfo {pages} {1} (\bibinfo {year} {2020})}\BibitemShut {NoStop}%
	\bibitem [{\citenamefont {Snider}\ \emph {et~al.}(2020)\citenamefont {Snider},
		\citenamefont {Dasenbrock-Gammon}, \citenamefont {McBride}, \citenamefont
		{Debessai}, \citenamefont {Vindana}, \citenamefont {Vencatasamy},
		\citenamefont {Lawler}, \citenamefont {Salamat},\ and\ \citenamefont
		{Dias}}]{Dias_Nature_2020_CSH}%
	\BibitemOpen
	\bibfield  {author} {\bibinfo {author} {\bibfnamefont {E.}~\bibnamefont
			{Snider}}, \bibinfo {author} {\bibfnamefont {N.}~\bibnamefont
			{Dasenbrock-Gammon}}, \bibinfo {author} {\bibfnamefont {R.}~\bibnamefont
			{McBride}}, \bibinfo {author} {\bibfnamefont {M.}~\bibnamefont {Debessai}},
		\bibinfo {author} {\bibfnamefont {H.}~\bibnamefont {Vindana}}, \bibinfo
		{author} {\bibfnamefont {K.}~\bibnamefont {Vencatasamy}}, \bibinfo {author}
		{\bibfnamefont {K.~V.}\ \bibnamefont {Lawler}}, \bibinfo {author}
		{\bibfnamefont {A.}~\bibnamefont {Salamat}},\ and\ \bibinfo {author}
		{\bibfnamefont {R.~P.}\ \bibnamefont {Dias}},\ }\href@noop {} {\bibfield
		{journal} {\bibinfo  {journal} {Nature}\ }\textbf {\bibinfo {volume} {586}},\
		\bibinfo {pages} {373} (\bibinfo {year} {2020})}\BibitemShut {NoStop}%
	\bibitem [{\citenamefont {Flores-Livas}\ \emph {et~al.}(2016)\citenamefont
		{Flores-Livas}, \citenamefont {Amsler}, \citenamefont {Heil}, \citenamefont
		{Sanna}, \citenamefont {Boeri}, \citenamefont {Profeta}, \citenamefont
		{Wolverton}, \citenamefont {Goedecker},\ and\ \citenamefont
		{Gross}}]{flores_PRBR2016}%
	\BibitemOpen
	\bibfield  {author} {\bibinfo {author} {\bibfnamefont {J.~A.}\ \bibnamefont
			{Flores-Livas}}, \bibinfo {author} {\bibfnamefont {M.}~\bibnamefont
			{Amsler}}, \bibinfo {author} {\bibfnamefont {C.}~\bibnamefont {Heil}},
		\bibinfo {author} {\bibfnamefont {A.}~\bibnamefont {Sanna}}, \bibinfo
		{author} {\bibfnamefont {L.}~\bibnamefont {Boeri}}, \bibinfo {author}
		{\bibfnamefont {G.}~\bibnamefont {Profeta}}, \bibinfo {author} {\bibfnamefont
			{C.}~\bibnamefont {Wolverton}}, \bibinfo {author} {\bibfnamefont
			{S.}~\bibnamefont {Goedecker}},\ and\ \bibinfo {author} {\bibfnamefont
			{E.~K.~U.}\ \bibnamefont {Gross}},\ }\href
	{https://doi.org/10.1103/PhysRevB.93.020508(R)} {\bibfield  {journal} {\bibinfo
			{journal} {Phys. Rev. B}\ }\textbf {\bibinfo {volume} {93}},\ \bibinfo
		{pages} {020508} (\bibinfo {year} {2016})}\BibitemShut {NoStop}%
	\bibitem [{\citenamefont {Bernstein}\ \emph {et~al.}(2015)\citenamefont
		{Bernstein}, \citenamefont {Hellberg}, \citenamefont {Johannes},\ and\
		\citenamefont {Mazin}}]{Mazin_PRB_2015_SH3}%
	\BibitemOpen
	\bibfield  {author} {\bibinfo {author} {\bibfnamefont {N.}~\bibnamefont
			{Bernstein}}, \bibinfo {author} {\bibfnamefont {C.~S.}\ \bibnamefont
			{Hellberg}}, \bibinfo {author} {\bibfnamefont {M.~D.}\ \bibnamefont
			{Johannes}},\ and\ \bibinfo {author} {\bibfnamefont {I.~I.}\ \bibnamefont
			{Mazin}},\ }\href@noop {} {\bibfield  {journal} {\bibinfo  {journal} {Phys.
				Rev. B}\ }\textbf {\bibinfo {volume} {91}} (\bibinfo {year}
		{2015})}\BibitemShut {NoStop}%
	\bibitem [{\citenamefont {Heil}\ and\ \citenamefont
		{Boeri}(2015)}]{Heil_PRB_2015_bonding}%
	\BibitemOpen
	\bibfield  {author} {\bibinfo {author} {\bibfnamefont {C.}~\bibnamefont
			{Heil}}\ and\ \bibinfo {author} {\bibfnamefont {L.}~\bibnamefont {Boeri}},\
	}\href@noop {} {\bibfield  {journal} {\bibinfo  {journal} {Phys. Rev. B}\
		}\textbf {\bibinfo {volume} {92}},\ \bibinfo {pages} {060508(R)} (\bibinfo
		{year} {2015})}\BibitemShut {NoStop}%
	\bibitem [{\citenamefont {Wang}\ \emph {et~al.}(2012)\citenamefont {Wang},
		\citenamefont {Tse}, \citenamefont {Tanaka}, \citenamefont {Iitaka},\ and\
		\citenamefont {Ma}}]{Ma_PNAS_2012_CaH}%
	\BibitemOpen
	\bibfield  {author} {\bibinfo {author} {\bibfnamefont {H.}~\bibnamefont
			{Wang}}, \bibinfo {author} {\bibfnamefont {J.~S.}\ \bibnamefont {Tse}},
		\bibinfo {author} {\bibfnamefont {K.}~\bibnamefont {Tanaka}}, \bibinfo
		{author} {\bibfnamefont {T.}~\bibnamefont {Iitaka}},\ and\ \bibinfo {author}
		{\bibfnamefont {Y.}~\bibnamefont {Ma}},\ }\href@noop {} {\bibfield  {journal}
		{\bibinfo  {journal} {PNAS}\ }\textbf {\bibinfo {volume} {109}},\ \bibinfo
		{pages} {6463} (\bibinfo {year} {2012})}\BibitemShut {NoStop}%
	\bibitem [{\citenamefont {Peng}\ \emph {et~al.}(2017)\citenamefont {Peng},
		\citenamefont {Sun}, \citenamefont {Pickard}, \citenamefont {Needs},
		\citenamefont {Wu},\ and\ \citenamefont {Ma}}]{Ma_PRL_2017_ReH}%
	\BibitemOpen
	\bibfield  {author} {\bibinfo {author} {\bibfnamefont {F.}~\bibnamefont
			{Peng}}, \bibinfo {author} {\bibfnamefont {Y.}~\bibnamefont {Sun}}, \bibinfo
		{author} {\bibfnamefont {C.~J.}\ \bibnamefont {Pickard}}, \bibinfo {author}
		{\bibfnamefont {R.~J.}\ \bibnamefont {Needs}}, \bibinfo {author}
		{\bibfnamefont {Q.}~\bibnamefont {Wu}},\ and\ \bibinfo {author}
		{\bibfnamefont {Y.}~\bibnamefont {Ma}},\ }\href@noop {} {\bibfield  {journal}
		{\bibinfo  {journal} {Phys. Rev. Lett.}\ }\textbf {\bibinfo {volume} {119}},\
		\bibinfo {pages} {107001} (\bibinfo {year} {2017})}\BibitemShut {NoStop}%
	\bibitem [{\citenamefont {Heil}\ \emph {et~al.}(2019)\citenamefont {Heil},
		\citenamefont {Cataldo}, \citenamefont {Bachelet},\ and\ \citenamefont
		{Boeri}}]{Heil_PRB_2019}%
	\BibitemOpen
	\bibfield  {author} {\bibinfo {author} {\bibfnamefont {C.}~\bibnamefont
			{Heil}}, \bibinfo {author} {\bibfnamefont {S.}\ \bibnamefont {Di Cataldo}},
		\bibinfo {author} {\bibfnamefont {G.~B.}\ \bibnamefont {Bachelet}},\ and\
		\bibinfo {author} {\bibfnamefont {L.}~\bibnamefont {Boeri}},\ }\href@noop {}
	{\bibfield  {journal} {\bibinfo  {journal} {Phys. Rev. B}\ }\textbf {\bibinfo
			{volume} {99}},\ \bibinfo {pages} {220502(R)} (\bibinfo {year}
		{2019})}\BibitemShut {NoStop}%
	\bibitem [{\citenamefont {Sun}\ and\ \citenamefont
		{Miao}(2021)}]{Miao_RS_2021_ChemTemplate}%
	\BibitemOpen
	\bibfield  {author} {\bibinfo {author} {\bibfnamefont {Y.}~\bibnamefont
			{Sun}}\ and\ \bibinfo {author} {\bibfnamefont {M.}~\bibnamefont {Miao}},\
	}\href@noop {} {\bibfield  {journal} {\bibinfo  {journal} {Preprint available
				(v1) at Research Square 10.21203/rs.3.rs-130093/v1}\ } (\bibinfo {year}
		{2021})}\BibitemShut {NoStop}%
	\bibitem [{\citenamefont {Yi}\ \emph {et~al.}(2021)\citenamefont {Yi},
		\citenamefont {Wang}, \citenamefont {Jeon},\ and\ \citenamefont
		{Cho}}]{Yi_PRM_2021_LaH10}%
	\BibitemOpen
	\bibfield  {author} {\bibinfo {author} {\bibfnamefont {S.}~\bibnamefont
			{Yi}}, \bibinfo {author} {\bibfnamefont {C.}~\bibnamefont {Wang}}, \bibinfo
		{author} {\bibfnamefont {H.}~\bibnamefont {Jeon}},\ and\ \bibinfo {author}
		{\bibfnamefont {J.-H.}\ \bibnamefont {Cho}},\ }\href@noop {} {\bibfield
		{journal} {\bibinfo  {journal} {Phys. Rev. M}\ }\textbf {\bibinfo {volume}
			{5}},\ \bibinfo {pages} {024801} (\bibinfo {year} {2021})}\BibitemShut
	{NoStop}%
	\bibitem [{\citenamefont {Song}\ \emph {et~al.}(2020)\citenamefont {Song},
		\citenamefont {Zhang}, \citenamefont {Cui2}, \citenamefont {Pickard},
		\citenamefont {Kresin},\ and\ \citenamefont {Duan}}]{Duan_arXiv_2020_ReH}%
	\BibitemOpen
	\bibfield  {author} {\bibinfo {author} {\bibfnamefont {H.}~\bibnamefont
			{Song}}, \bibinfo {author} {\bibfnamefont {Z.}~\bibnamefont {Zhang}},
		\bibinfo {author} {\bibfnamefont {T.}~\bibnamefont {Cui2}}, \bibinfo {author}
		{\bibfnamefont {C.~J.}\ \bibnamefont {Pickard}}, \bibinfo {author}
		{\bibfnamefont {V.~Z.}\ \bibnamefont {Kresin}},\ and\ \bibinfo {author}
		{\bibfnamefont {D.}~\bibnamefont {Duan}},\ }\href@noop {} {\bibfield
		{journal} {\bibinfo  {journal} {arXiv:2010.12225}\ } (\bibinfo {year}
		{2020})}\BibitemShut {NoStop}%
	\bibitem [{\citenamefont {Guigue}\ \emph {et~al.}(2020)\citenamefont {Guigue},
		\citenamefont {Marizy},\ and\ \citenamefont
		{Loubeyre}}]{Guigue_PRB_2020_UH7}%
	\BibitemOpen
	\bibfield  {author} {\bibinfo {author} {\bibfnamefont {B.}~\bibnamefont
			{Guigue}}, \bibinfo {author} {\bibfnamefont {A.}~\bibnamefont {Marizy}},\
		and\ \bibinfo {author} {\bibfnamefont {P.}~\bibnamefont {Loubeyre}},\
	}\href@noop {} {\bibfield  {journal} {\bibinfo  {journal} {Phys. Rev. B}\
		}\textbf {\bibinfo {volume} {102}},\ \bibinfo {pages} {014107} (\bibinfo
		{year} {2020})}\BibitemShut {NoStop}%
	\bibitem [{\citenamefont {Kokail}\ \emph {et~al.}(2017)\citenamefont {Kokail},
		\citenamefont {von~der Linden},\ and\ \citenamefont
		{Boeri}}]{Kokail_PRM_2017_LiBH}%
	\BibitemOpen
	\bibfield  {author} {\bibinfo {author} {\bibfnamefont {C.}~\bibnamefont
			{Kokail}}, \bibinfo {author} {\bibfnamefont {W.}~\bibnamefont {von~der
				Linden}},\ and\ \bibinfo {author} {\bibfnamefont {L.}~\bibnamefont {Boeri}},\
	}\href@noop {} {\bibfield  {journal} {\bibinfo  {journal} {Phys. Rev. M}\
		}\textbf {\bibinfo {volume} {1}},\ \bibinfo {pages} {074803} (\bibinfo {year}
		{2017})}\BibitemShut {NoStop}%
	\bibitem [{\citenamefont {Cataldo}\ \emph {et~al.}(2020)\citenamefont
		{Cataldo}, \citenamefont {von~der Linden},\ and\ \citenamefont
		{Boeri}}]{DiCataldo_PRB_2020_CaBH}%
	\BibitemOpen
	\bibfield  {author} {\bibinfo {author} {\bibfnamefont {S.}\ \bibnamefont
			{Di Cataldo}}, \bibinfo {author} {\bibfnamefont {W.}~\bibnamefont {von~der
				Linden}},\ and\ \bibinfo {author} {\bibfnamefont {L.}~\bibnamefont {Boeri}},\
	}\href@noop {} {\bibfield  {journal} {\bibinfo  {journal} {Phys. Rev. B}\
		}\textbf {\bibinfo {volume} {102}},\ \bibinfo {pages} {014516} (\bibinfo
		{year} {2020})}\BibitemShut {NoStop}%
	\bibitem [{\citenamefont {Sun}\ \emph {et~al.}(2019)\citenamefont {Sun},
		\citenamefont {Lv}, \citenamefont {Xie}, \citenamefont {Liu},\ and\
		\citenamefont {Ma}}]{Ma_PRL_2019_Li2MgH16}%
	\BibitemOpen
	\bibfield  {author} {\bibinfo {author} {\bibfnamefont {Y.}~\bibnamefont
			{Sun}}, \bibinfo {author} {\bibfnamefont {J.}~\bibnamefont {Lv}}, \bibinfo
		{author} {\bibfnamefont {Y.}~\bibnamefont {Xie}}, \bibinfo {author}
		{\bibfnamefont {H.}~\bibnamefont {Liu}},\ and\ \bibinfo {author}
		{\bibfnamefont {Y.}~\bibnamefont {Ma}},\ }\href@noop {} {\bibfield  {journal}
		{\bibinfo  {journal} {Phys. Rev. Lett.}\ }\textbf {\bibinfo {volume} {123}},\
		\bibinfo {pages} {097001} (\bibinfo {year} {2019})}\BibitemShut {NoStop}%
	\bibitem [{\citenamefont {Semenok}\ \emph
		{et~al.}(2020{\natexlab{b}})\citenamefont {Semenok}, \citenamefont {Troyan},
		\citenamefont {Kvashin}, \citenamefont {Ivanova}, \citenamefont {Hanfland},
		\citenamefont {Sadakov}, \citenamefont {Sobolevskiy}, \citenamefont
		{Pervakov}, \citenamefont {Gavriliuk}, \citenamefont {Lyubutin},
		\citenamefont {Glazyrin}, \citenamefont {Giordano}, \citenamefont {Karimov},
		\citenamefont {Vasiliev}, \citenamefont {Akashi}, \citenamefont {Pudalov},\
		and\ \citenamefont {Oganov}}]{Oganov_arXiv_LaYH_2020}%
	\BibitemOpen
	\bibfield  {author} {\bibinfo {author} {\bibfnamefont {D.~V.}\ \bibnamefont
			{Semenok}}, \bibinfo {author} {\bibfnamefont {I.~A.}\ \bibnamefont {Troyan}},
		\bibinfo {author} {\bibfnamefont {A.~G.}\ \bibnamefont {Kvashin}}, \bibinfo
		{author} {\bibfnamefont {A.~G.}\ \bibnamefont {Ivanova}}, \bibinfo {author}
		{\bibfnamefont {M.}~\bibnamefont {Hanfland}}, \bibinfo {author}
		{\bibfnamefont {A.~V.}\ \bibnamefont {Sadakov}}, \bibinfo {author}
		{\bibfnamefont {O.~A.}\ \bibnamefont {Sobolevskiy}}, \bibinfo {author}
		{\bibfnamefont {K.~S.}\ \bibnamefont {Pervakov}}, \bibinfo {author}
		{\bibfnamefont {A.~G.}\ \bibnamefont {Gavriliuk}}, \bibinfo {author}
		{\bibfnamefont {I.~S.}\ \bibnamefont {Lyubutin}}, \bibinfo {author}
		{\bibfnamefont {K.~V.}\ \bibnamefont {Glazyrin}}, \bibinfo {author}
		{\bibfnamefont {N.}~\bibnamefont {Giordano}}, \bibinfo {author}
		{\bibfnamefont {D.~N.}\ \bibnamefont {Karimov}}, \bibinfo {author}
		{\bibfnamefont {A.~B.}\ \bibnamefont {Vasiliev}}, \bibinfo {author}
		{\bibfnamefont {R.}~\bibnamefont {Akashi}}, \bibinfo {author} {\bibfnamefont
			{V.~M.}\ \bibnamefont {Pudalov}},\ and\ \bibinfo {author} {\bibfnamefont
			{A.~R.}\ \bibnamefont {Oganov}},\ }\href@noop {} {\bibfield  {journal}
		{\bibinfo  {journal} {arXiv:2012.04787}\ } (\bibinfo {year}
		{2020}{\natexlab{b}})}\BibitemShut {NoStop}%
	\bibitem [{\citenamefont {Glass}\ \emph {et~al.}(2006)\citenamefont {Glass},
		\citenamefont {Oganov},\ and\ \citenamefont {Hansen}}]{USPEX_1}%
	\BibitemOpen
	\bibfield  {author} {\bibinfo {author} {\bibfnamefont {C.~W.}\ \bibnamefont
			{Glass}}, \bibinfo {author} {\bibfnamefont {A.~R.}\ \bibnamefont {Oganov}},\
		and\ \bibinfo {author} {\bibfnamefont {N.}~\bibnamefont {Hansen}},\
	}\href@noop {} {\bibfield  {journal} {\bibinfo  {journal} {Computer Physics
				Communication}\ }\textbf {\bibinfo {volume} {175}},\ \bibinfo {pages} {713}
		(\bibinfo {year} {2006})}\BibitemShut {NoStop}%
	\bibitem [{\citenamefont {Lyakhov}\ \emph {et~al.}(2013)\citenamefont
		{Lyakhov}, \citenamefont {Oganov}, \citenamefont {Stokes},\ and\
		\citenamefont {Zhu}}]{USPEX_2}%
	\BibitemOpen
	\bibfield  {author} {\bibinfo {author} {\bibfnamefont {A.~O.}\ \bibnamefont
			{Lyakhov}}, \bibinfo {author} {\bibfnamefont {A.~R.}\ \bibnamefont {Oganov}},
		\bibinfo {author} {\bibfnamefont {H.~T.}\ \bibnamefont {Stokes}},\ and\
		\bibinfo {author} {\bibfnamefont {Q.}~\bibnamefont {Zhu}},\ }\href@noop {}
	{\bibfield  {journal} {\bibinfo  {journal} {Computer Physics Communication}\
		}\textbf {\bibinfo {volume} {184}},\ \bibinfo {pages} {1172} (\bibinfo {year}
		{2013})}\BibitemShut {NoStop}%
	\bibitem [{\citenamefont {Liang}\ \emph {et~al.}(2019)\citenamefont {Liang},
		\citenamefont {Bergara}, \citenamefont {Wang}, \citenamefont {Wen},
		\citenamefont {Zhao}, \citenamefont {Zhou}, \citenamefont {He}, \citenamefont
		{Gao},\ and\ \citenamefont {Tian}}]{Liang_PRB_2019_CaYH}%
	\BibitemOpen
	\bibfield  {author} {\bibinfo {author} {\bibfnamefont {X.}~\bibnamefont
			{Liang}}, \bibinfo {author} {\bibfnamefont {A.}~\bibnamefont {Bergara}},
		\bibinfo {author} {\bibfnamefont {L.}~\bibnamefont {Wang}}, \bibinfo {author}
		{\bibfnamefont {B.}~\bibnamefont {Wen}}, \bibinfo {author} {\bibfnamefont
			{Z.}~\bibnamefont {Zhao}}, \bibinfo {author} {\bibfnamefont {X.-F.}\
			\bibnamefont {Zhou}}, \bibinfo {author} {\bibfnamefont {J.}~\bibnamefont
			{He}}, \bibinfo {author} {\bibfnamefont {G.}~\bibnamefont {Gao}},\ and\
		\bibinfo {author} {\bibfnamefont {Y.}~\bibnamefont {Tian}},\ }\href@noop {}
	{\bibfield  {journal} {\bibinfo  {journal} {Phys. Rev. B}\ }\textbf {\bibinfo
			{volume} {99}},\ \bibinfo {pages} {100505(R)} (\bibinfo {year}
		{2019})}\BibitemShut {NoStop}%
	\bibitem [{Note1()}]{Note1}%
	\BibitemOpen
	\bibinfo {note} {In addition, we re-sampled particularly promising
		compositions. Further details are provided in the Supplemental
		Material \cite{Note3}}\BibitemShut {NoStop}%
	\bibitem [{\citenamefont {Paskevicius}\ \emph {et~al.}(2017)\citenamefont
		{Paskevicius}, \citenamefont {Jepsen}, \citenamefont {Schouwink},
		\citenamefont {Cern\'{y}}, \citenamefont {Ravnsbaek}, \citenamefont
		{Filinchuk}, \citenamefont {Dornheim}, \citenamefont {Besenbacher},\ and\
		\citenamefont {Jensen}}]{Jensen_ChemSocRev_2017_MBH}%
	\BibitemOpen
	\bibfield  {author} {\bibinfo {author} {\bibfnamefont {M.}~\bibnamefont
			{Paskevicius}}, \bibinfo {author} {\bibfnamefont {L.~H.}\ \bibnamefont
			{Jepsen}}, \bibinfo {author} {\bibfnamefont {P.}~\bibnamefont {Schouwink}},
		\bibinfo {author} {\bibfnamefont {R.}~\bibnamefont {Cern\'{y}}}, \bibinfo
		{author} {\bibfnamefont {D.~B.}\ \bibnamefont {Ravnsbaek}}, \bibinfo {author}
		{\bibfnamefont {Y.}~\bibnamefont {Filinchuk}}, \bibinfo {author}
		{\bibfnamefont {M.}~\bibnamefont {Dornheim}}, \bibinfo {author}
		{\bibfnamefont {F.}~\bibnamefont {Besenbacher}},\ and\ \bibinfo {author}
		{\bibfnamefont {T.~R.}\ \bibnamefont {Jensen}},\ }\href@noop {} {\bibfield
		{journal} {\bibinfo  {journal} {Chem. Soc. Rev.}\ }\textbf {\bibinfo {volume}
			{46}},\ \bibinfo {pages} {1565} (\bibinfo {year} {2017})}\BibitemShut
	{NoStop}%
	\bibitem [{\citenamefont {Zhang}\ \emph {et~al.}(2010)\citenamefont {Zhang},
		\citenamefont {Majzoub.}, \citenamefont {Ozolins},\ and\ \citenamefont
		{Wolverton}}]{Wolverton_PRB_2010_CaBHx}%
	\BibitemOpen
	\bibfield  {author} {\bibinfo {author} {\bibfnamefont {Y.}~\bibnamefont
			{Zhang}}, \bibinfo {author} {\bibfnamefont {E.}~\bibnamefont {Majzoub.}},
		\bibinfo {author} {\bibfnamefont {V.}~\bibnamefont {Ozolins}},\ and\ \bibinfo
		{author} {\bibfnamefont {C.}~\bibnamefont {Wolverton}},\ }\href@noop {}
	{\bibfield  {journal} {\bibinfo  {journal} {Phys. Rev. B}\ }\textbf {\bibinfo
			{volume} {82}},\ \bibinfo {pages} {174107} (\bibinfo {year}
		{2010})}\BibitemShut {NoStop}%
	\bibitem [{\citenamefont {Grinderslev}\ \emph {et~al.}(2020)\citenamefont
		{Grinderslev}, \citenamefont {Ley}, \citenamefont {Lee}, \citenamefont
		{Jepsen}, \citenamefont {rgensen}, \citenamefont {Cho}, \citenamefont {rgen
			Skibsted},\ and\ \citenamefont {Jensen}}]{Jensen_IC_2020_LaB3H12}%
	\BibitemOpen
	\bibfield  {author} {\bibinfo {author} {\bibfnamefont {J.~B.}\ \bibnamefont
			{Grinderslev}}, \bibinfo {author} {\bibfnamefont {M.~B.}\ \bibnamefont
			{Ley}}, \bibinfo {author} {\bibfnamefont {Y.-S.}\ \bibnamefont {Lee}},
		\bibinfo {author} {\bibfnamefont {L.~H.}\ \bibnamefont {Jepsen}}, \bibinfo
		{author} {\bibfnamefont {M.~J.}\ \bibnamefont {rgensen}}, \bibinfo {author}
		{\bibfnamefont {Y.~W.}\ \bibnamefont {Cho}}, \bibinfo {author} {\bibfnamefont
			{J.}~\bibnamefont {rgen Skibsted}},\ and\ \bibinfo {author} {\bibfnamefont
			{T.~R.}\ \bibnamefont {Jensen}},\ }\href@noop {} {\bibfield  {journal}
		{\bibinfo  {journal} {Inorganic Chemistry}\ }\textbf {\bibinfo {volume}
			{59}},\ \bibinfo {pages} {7768} (\bibinfo {year} {2020})}\BibitemShut
	{NoStop}%
	\bibitem [{\citenamefont {Rude}\ \emph {et~al.}(2011)\citenamefont {Rude},
		\citenamefont {Nielsen}, \citenamefont {Ravnsbaek}, \citenamefont
		{B\"{o}senberg}, \citenamefont {Ley}, \citenamefont {Richter}, \citenamefont
		{Arnbjerg}, \citenamefont {Dornheim}, \citenamefont {Filinchuk},
		\citenamefont {Besenbacher},\ and\ \citenamefont
		{Jensen}}]{Jensen_PSS_2011_MBH}%
	\BibitemOpen
	\bibfield  {author} {\bibinfo {author} {\bibfnamefont {L.~H.}\ \bibnamefont
			{Rude}}, \bibinfo {author} {\bibfnamefont {T.~K.}\ \bibnamefont {Nielsen}},
		\bibinfo {author} {\bibfnamefont {D.~B.}\ \bibnamefont {Ravnsbaek}}, \bibinfo
		{author} {\bibfnamefont {U.}~\bibnamefont {B\"{o}senberg}}, \bibinfo {author}
		{\bibfnamefont {M.~B.}\ \bibnamefont {Ley}}, \bibinfo {author} {\bibfnamefont
			{B.}~\bibnamefont {Richter}}, \bibinfo {author} {\bibfnamefont {L.~M.}\
			\bibnamefont {Arnbjerg}}, \bibinfo {author} {\bibfnamefont {M.}~\bibnamefont
			{Dornheim}}, \bibinfo {author} {\bibfnamefont {Y.}~\bibnamefont {Filinchuk}},
		\bibinfo {author} {\bibfnamefont {F.}~\bibnamefont {Besenbacher}},\ and\
		\bibinfo {author} {\bibfnamefont {T.~R.}\ \bibnamefont {Jensen}},\
	}\href@noop {} {\bibfield  {journal} {\bibinfo  {journal} {Phys. Status
				Solidi}\ }\textbf {\bibinfo {volume} {208}},\ \bibinfo {pages} {1754}
		(\bibinfo {year} {2011})}\BibitemShut {NoStop}%
	\bibitem [{Note2()}]{Note2}%
	\BibitemOpen
	\bibinfo {note} {Additional information on the crystal structures can be
		found in the form of Crystallographic Information File in the Supplemental
		Material \cite{Note3}}\BibitemShut {NoStop}%
	\bibitem [{\citenamefont {Renaudin}\ \emph {et~al.}(2004)\citenamefont
		{Renaudin}, \citenamefont {Gomes}, \citenamefont {Hagemann}, \citenamefont
		{Keller},\ and\ \citenamefont {Yvon}}]{Yvon_JAC_2004_structure_CsBH4}%
	\BibitemOpen
	\bibfield  {author} {\bibinfo {author} {\bibfnamefont {G.}~\bibnamefont
			{Renaudin}}, \bibinfo {author} {\bibfnamefont {S.}~\bibnamefont {Gomes}},
		\bibinfo {author} {\bibfnamefont {H.}~\bibnamefont {Hagemann}}, \bibinfo
		{author} {\bibfnamefont {L.}~\bibnamefont {Keller}},\ and\ \bibinfo {author}
		{\bibfnamefont {K.}~\bibnamefont {Yvon}},\ }\href@noop {} {\bibfield
		{journal} {\bibinfo  {journal} {Journal of Alloys and Compounds}\ }\textbf
		{\bibinfo {volume} {375}},\ \bibinfo {pages} {98} (\bibinfo {year}
		{2004})}\BibitemShut {NoStop}%
	\bibitem [{Note3()}]{Note3}%
	\BibitemOpen
	\bibinfo {note} {The Supplemental Material is available at..}\BibitemShut
	{Stop}%
	\bibitem [{\citenamefont {Sanville}\ \emph {et~al.}(2007)\citenamefont
		{Sanville}, \citenamefont {Kenny}, \citenamefont {Smith},\ and\ \citenamefont
		{Henkelmann}}]{Henkelman_JCC_2007_Bader}%
	\BibitemOpen
	\bibfield  {author} {\bibinfo {author} {\bibfnamefont {E.}~\bibnamefont
			{Sanville}}, \bibinfo {author} {\bibfnamefont {S.~D.}\ \bibnamefont {Kenny}},
		\bibinfo {author} {\bibfnamefont {R.}~\bibnamefont {Smith}},\ and\ \bibinfo
		{author} {\bibfnamefont {G.}~\bibnamefont {Henkelmann}},\ }\href@noop {}
	{\bibfield  {journal} {\bibinfo  {journal} {J. Comp. Chem.}\ }\textbf
		{\bibinfo {volume} {28}},\ \bibinfo {pages} {899} (\bibinfo {year}
		{2007})}\BibitemShut {NoStop}%
	\bibitem [{\citenamefont {Baroni}\ \emph {et~al.}(2001)\citenamefont {Baroni},
		\citenamefont {de~Gironcoli}, \citenamefont {Corso},\ and\ \citenamefont
		{Giannozzi}}]{Baroni_RevModPhys_2001_DFPT}%
	\BibitemOpen
	\bibfield  {author} {\bibinfo {author} {\bibfnamefont {S.}~\bibnamefont
			{Baroni}}, \bibinfo {author} {\bibfnamefont {S.}~\bibnamefont
			{de~Gironcoli}}, \bibinfo {author} {\bibfnamefont {A.~D.}\ \bibnamefont
			{Corso}},\ and\ \bibinfo {author} {\bibfnamefont {P.}~\bibnamefont
			{Giannozzi}},\ }\href@noop {} {\bibfield  {journal} {\bibinfo  {journal}
			{Rev. Mod. Phys}\ }\textbf {\bibinfo {volume} {73}},\ \bibinfo {pages} {515}
		(\bibinfo {year} {2001})}\BibitemShut {NoStop}%
	\bibitem [{\citenamefont {Savrasov}\ and\ \citenamefont
		{Savrasov}(1990)}]{Savrasov_PRB_1996_lrt}%
	\BibitemOpen
	\bibfield  {author} {\bibinfo {author} {\bibfnamefont {S.~Y.}\ \bibnamefont
			{Savrasov}}\ and\ \bibinfo {author} {\bibfnamefont {D.~Y.}\ \bibnamefont
			{Savrasov}},\ }\href@noop {} {\bibfield  {journal} {\bibinfo  {journal}
			{Phys. Rev. B}\ }\textbf {\bibinfo {volume} {54}},\ \bibinfo {pages} {16487}
		(\bibinfo {year} {1996})}\BibitemShut {NoStop}%
	\bibitem [{\citenamefont {and. E.~R.~Margine}\ \emph
		{et~al.}(2016)\citenamefont {and. E.~R.~Margine}, \citenamefont {Verdi},\
		and\ \citenamefont {Giustino}}]{Giustino_CPC_2016_EPW}%
	\BibitemOpen
	\bibfield  {author} {\bibinfo {author} {\bibfnamefont {S.~P.}\ \bibnamefont
			{and. E.~R.~Margine}}, \bibinfo {author} {\bibfnamefont {C.}~\bibnamefont
			{Verdi}},\ and\ \bibinfo {author} {\bibfnamefont {F.}~\bibnamefont
			{Giustino}},\ }\href@noop {} {\bibfield  {journal} {\bibinfo  {journal}
			{Comp. Phys. Communications}\ }\textbf {\bibinfo {volume} {209}},\ \bibinfo
		{pages} {116} (\bibinfo {year} {2016})}\BibitemShut {NoStop}%
	\bibitem [{\citenamefont {Errea}\ \emph {et~al.}(2020)\citenamefont {Errea},
		\citenamefont {Belli}, \citenamefont {Monacelli}, \citenamefont {Sanna},
		\citenamefont {Koretsune}, \citenamefont {Tadano}, \citenamefont {Bianco},
		\citenamefont {Calandra}, \citenamefont {Arita}, \citenamefont {Mauri},\ and\
		\citenamefont {Flores-Livas}}]{Mauri_Nature_2020_LaH}%
	\BibitemOpen
	\bibfield  {author} {\bibinfo {author} {\bibfnamefont {I.}~\bibnamefont
			{Errea}}, \bibinfo {author} {\bibfnamefont {F.}~\bibnamefont {Belli}},
		\bibinfo {author} {\bibfnamefont {L.}~\bibnamefont {Monacelli}}, \bibinfo
		{author} {\bibfnamefont {A.}~\bibnamefont {Sanna}}, \bibinfo {author}
		{\bibfnamefont {T.}~\bibnamefont {Koretsune}}, \bibinfo {author}
		{\bibfnamefont {T.}~\bibnamefont {Tadano}}, \bibinfo {author} {\bibfnamefont
			{R.}~\bibnamefont {Bianco}}, \bibinfo {author} {\bibfnamefont
			{M.}~\bibnamefont {Calandra}}, \bibinfo {author} {\bibfnamefont
			{R.}~\bibnamefont {Arita}}, \bibinfo {author} {\bibfnamefont
			{F.}~\bibnamefont {Mauri}},\ and\ \bibinfo {author} {\bibfnamefont {J.~A.}\
			\bibnamefont {Flores-Livas}},\ }\href@noop {} {\bibfield  {journal} {\bibinfo
			{journal} {Nature}\ }\textbf {\bibinfo {volume} {578}},\ \bibinfo {pages}
		{66} (\bibinfo {year} {2020})}\BibitemShut {NoStop}%
	\bibitem [{\citenamefont {McMillan}(1968)}]{McMillan_PR_1968}%
	\BibitemOpen
	\bibfield  {author} {\bibinfo {author} {\bibfnamefont {W.~L.}\ \bibnamefont
			{McMillan}},\ }\href@noop {} {\bibfield  {journal} {\bibinfo  {journal}
			{Physical Review}\ }\textbf {\bibinfo {volume} {167}},\ \bibinfo {pages}
		{331} (\bibinfo {year} {1968})}\BibitemShut {NoStop}%
	\bibitem [{\citenamefont {Allen}\ and\ \citenamefont
		{Dynes}(1975)}]{Allen_PRB_1975_McMillan}%
	\BibitemOpen
	\bibfield  {author} {\bibinfo {author} {\bibfnamefont {P.~B.}\ \bibnamefont
			{Allen}}\ and\ \bibinfo {author} {\bibfnamefont {R.~C.}\ \bibnamefont
			{Dynes}},\ }\href@noop {} {\bibfield  {journal} {\bibinfo  {journal} {Phys.
				Rev. B}\ }\textbf {\bibinfo {volume} {12}},\ \bibinfo {pages} {905} (\bibinfo
		{year} {1975})}\BibitemShut {NoStop}%
	\bibitem [{\citenamefont {Morel}\ and\ \citenamefont
		{Anderson}(1962)}]{Morel_PhysRev_1962_mustar}%
	\BibitemOpen
	\bibfield  {author} {\bibinfo {author} {\bibfnamefont {P.}~\bibnamefont
			{Morel}}\ and\ \bibinfo {author} {\bibfnamefont {P.~W.}\ \bibnamefont
			{Anderson}},\ }\href@noop {} {\bibfield  {journal} {\bibinfo  {journal}
			{Physical Review}\ }\textbf {\bibinfo {volume} {125}},\ \bibinfo {pages}
		{1263} (\bibinfo {year} {1962})}\BibitemShut {NoStop}%
	\bibitem [{\citenamefont {Lee}\ \emph {et~al.}(1995)\citenamefont {Lee},
		\citenamefont {Chang},\ and\ \citenamefont {Cohen}}]{Lee_PRB_1995_mustar}%
	\BibitemOpen
	\bibfield  {author} {\bibinfo {author} {\bibfnamefont {K.-H.}\ \bibnamefont
			{Lee}}, \bibinfo {author} {\bibfnamefont {K.~J.}\ \bibnamefont {Chang}},\
		and\ \bibinfo {author} {\bibfnamefont {M.~L.}\ \bibnamefont {Cohen}},\
	}\href@noop {} {\bibfield  {journal} {\bibinfo  {journal} {Phys. Rev. B}\
		}\textbf {\bibinfo {volume} {52}},\ \bibinfo {pages} {1425} (\bibinfo {year}
		{1995})}\BibitemShut {NoStop}%
	\bibitem [{\citenamefont {Giustino}\ \emph {et~al.}(2010)\citenamefont
		{Giustino}, \citenamefont {Cohen},\ and\ \citenamefont
		{Louie}}]{Giustino_PRB_2010_GWsternheimer}%
	\BibitemOpen
	\bibfield  {author} {\bibinfo {author} {\bibfnamefont {F.}~\bibnamefont
			{Giustino}}, \bibinfo {author} {\bibfnamefont {M.~L.}\ \bibnamefont
			{Cohen}},\ and\ \bibinfo {author} {\bibfnamefont {S.~G.}\ \bibnamefont
			{Louie}},\ }\href@noop {} {\bibfield  {journal} {\bibinfo  {journal} {Phys.
				Rev. B}\ }\textbf {\bibinfo {volume} {81}},\ \bibinfo {pages} {115105}
		(\bibinfo {year} {2010})}\BibitemShut {NoStop}%
	\bibitem [{\citenamefont {Lambert}\ and\ \citenamefont
		{Giustino}(2013)}]{Lambert_PRB_2013_GWsternheimer}%
	\BibitemOpen
	\bibfield  {author} {\bibinfo {author} {\bibfnamefont {H.}~\bibnamefont
			{Lambert}}\ and\ \bibinfo {author} {\bibfnamefont {F.}~\bibnamefont
			{Giustino}},\ }\href@noop {} {\bibfield  {journal} {\bibinfo  {journal}
			{Phys. Rev. B}\ }\textbf {\bibinfo {volume} {88}},\ \bibinfo {pages} {075117}
		(\bibinfo {year} {2013})}\BibitemShut {NoStop}%
	\bibitem [{\citenamefont {Heil}\ \emph {et~al.}(2017)\citenamefont {Heil},
		\citenamefont {Ponc\'{e}}, \citenamefont {Lambert}, \citenamefont {Schlipf},
		\citenamefont {Margine},\ and\ \citenamefont
		{Giustino}}]{Heil_PRL_2017_NbS2}%
	\BibitemOpen
	\bibfield  {author} {\bibinfo {author} {\bibfnamefont {C.}~\bibnamefont
			{Heil}}, \bibinfo {author} {\bibfnamefont {S.}~\bibnamefont {Ponc\'{e}}},
		\bibinfo {author} {\bibfnamefont {H.}~\bibnamefont {Lambert}}, \bibinfo
		{author} {\bibfnamefont {M.}~\bibnamefont {Schlipf}}, \bibinfo {author}
		{\bibfnamefont {E.~R.}\ \bibnamefont {Margine}},\ and\ \bibinfo {author}
		{\bibfnamefont {F.}~\bibnamefont {Giustino}},\ }\href@noop {} {\bibfield
		{journal} {\bibinfo  {journal} {Phys. Rev. Lett.}\ }\textbf {\bibinfo
			{volume} {119}},\ \bibinfo {pages} {087003} (\bibinfo {year}
		{2017})}\BibitemShut {NoStop}%
	\bibitem [{Note4()}]{Note4}%
	\BibitemOpen
	\bibinfo {note} {We also checked the dependence of $\mu ^{*}$ on Tc and found
		that a variation of 0.01 in $\mu ^{*}$ changes Tc only by 2-3 K (See SM Fig.
		S9 for more details \cite{Note3})}\BibitemShut {Stop}%
	\bibitem [{\citenamefont {Errea}\ \emph {et~al.}(2016)\citenamefont {Errea},
		\citenamefont {Calandra}, \citenamefont {Pickard}, \citenamefont {Nelson},
		\citenamefont {Needs}, \citenamefont {Li}, \citenamefont {Liu}, \citenamefont
		{Zhang}, \citenamefont {Ma},\ and\ \citenamefont
		{Mauri}}]{Mauri_Nature_2016_SH3}%
	\BibitemOpen
	\bibfield  {author} {\bibinfo {author} {\bibfnamefont {I.}~\bibnamefont
			{Errea}}, \bibinfo {author} {\bibfnamefont {M.}~\bibnamefont {Calandra}},
		\bibinfo {author} {\bibfnamefont {C.~J.}\ \bibnamefont {Pickard}}, \bibinfo
		{author} {\bibfnamefont {J.~R.}\ \bibnamefont {Nelson}}, \bibinfo {author}
		{\bibfnamefont {R.~J.}\ \bibnamefont {Needs}}, \bibinfo {author}
		{\bibfnamefont {Y.}~\bibnamefont {Li}}, \bibinfo {author} {\bibfnamefont
			{H.}~\bibnamefont {Liu}}, \bibinfo {author} {\bibfnamefont {Y.}~\bibnamefont
			{Zhang}}, \bibinfo {author} {\bibfnamefont {Y.}~\bibnamefont {Ma}},\ and\
		\bibinfo {author} {\bibfnamefont {F.}~\bibnamefont {Mauri}},\ }\href@noop {}
	{\bibfield  {journal} {\bibinfo  {journal} {Nature}\ }\textbf {\bibinfo
			{volume} {532}},\ \bibinfo {pages} {81} (\bibinfo {year} {2016})}\BibitemShut
	{NoStop}%
\end{thebibliography}

\end{document}